\documentclass{pasj00}
\draft

\begin{document}
\SetRunningHead{Author(s) in page-head}{Running Head}

\title{The Abundance Inhomogeneity in the Northern Rim of the Cygnus Loop}

\author{Hiroyuki \textsc{Uchida}\altaffilmark{1}, Hiroshi \textsc{Tsunemi}\altaffilmark{1}, Satoru \textsc{Katsuda}\altaffilmark{1,2}, Masashi \textsc{Kimura}\altaffilmark{1}, Hiroko \textsc{Kosugi}\altaffilmark{1}, Hiroaki \textsc{Takahashi}\altaffilmark{1}} %
\altaffiltext{1}{Department of Earth and Space Science, Graduate School of
  Science, Osaka University, Toyonaka, Osaka 560-0043, Japan}
\altaffiltext{2}{Code 662, NASA Goddard Space Flight Center, Greenbelt, MD 20771}
\email{uchida@ess.sci.osaka-u.ac.jp}


\maketitle

\begin{abstract}
We observed the northern rim of the Cygnus Loop with the \textit{Suzaku} observatory in 5 pointings (P21-P25).
From the spatially resolved analysis, all the spectra are well fitted by the single component of the non-equilibrium ionization plasma model. From the best-fit parameters, we found that the abundances of the heavy elements are significantly lower than the solar values except those at the outermost edge in P21 and P22.
The origin of the depleted metal abundances is still unclear while such deficiencies have been reported from many other rim observations of the Loop. 
To explain these depletion at the rim regions, we considered the several possibilities. 
The effects of the resonance-line-scattering and the grain condensation lower the values of the abundances.
However, these are not sufficient to account for the abundance depletion observed.

We found that the abundances at the outermost edge in P21 and P22 are higher than those at the other regions.
From the morphological point of view, it is reasonable to consider that this abundance inhomogeneity is derived from the breakout or the thinness of the cavity wall of the Loop.
\end{abstract}

\section{Introduction}\label{sec:intro}
The Cygnus Loop is one of the brightest supernova remnants (SNRs) in the X-ray sky.
The age is estimated to be $\sim$10,000 yr and the distance is comparatively close to us (540\,pc; \cite{Blair05}).
The apparent size is quite large ($2^\circ.5\times3^\circ$.5; \cite{Levenson97}), which enables us to study the plasma structure of the Loop.
From the morphological point of view, the Cygnus Loop is a typical shell-like SNR and it is almost circular in shape. 
This structure is thought to be generated by a cavity explosion \citep{Levenson97}.

Although the Cygnus Loop is an evolved SNR, a hot plasma is still confined inside the Loop (\cite{Tsunemi88}; \cite{Hatsukade90}). 
\citet{Tsunemi07} observed the Cygnus Loop along the diameter from the northeast (NE) to the southwest (SW) with the \textit{XMM-Newton} observatory and showed that the Cygnus Loop consists of two components with different temperatures and metal abundances. 
They concluded that the low-$kT_e$ component originating from the cavity wall surrounds the high-$kT_e$ ejecta component. 
From the metal abundances of the ejecta component, they calculated the progenitor mass of the Cygnus Loop and concluded that this SNR is originated from the 12-15\MO \ explosion. 
\citet{Levenson98} also estimated the progenitor mass to be 15\MO \ from the size of the cavity. 
These results show that the progenitor star of the Cygnus Loop was a massive star and caused a core-collapse explosion.

\citet{Miyata07} observed the NE rim (NE2) of the Loop with the \textit{Suzaku} observatory and showed the abundances of C to Fe to be depleted (typically $\sim$0.1 times solar) in their field of view (FOV).
\citet{Katsuda08NE} expanded their observation northward (NE1-NE4; including the FOV of \cite{Miyata07}) with \textit{Suzaku} and found that a portion of their FOV, the outer edge of the rim in NE3-NE4, only shows the relatively high abundances while the other regions are uniformly depleted.
\citet{Katsuda08Chan} confirmed it by \textit{Chandra}.
\citet{Tsunemi09}\footnote{The data available at http://arxiv.org/pdf/0810.5209v1} also found the abundance-enhanced region at the southeastern (SE) rim with the \textit{Suzaku} observatory.
Although \citet{Katsuda08Chan} and \citet{Tsunemi09} discussed a few possibilities for the origin of the abundance inhomogeneities, the origin of these inhomogeneities remains in question.

We observed the northern rim of the Cygnus Loop with \textit{Suzaku} in 5 pointings.
Our region observed is contiguous with that of \citet{Katsuda08NE}.
We report here the result of the observation showing the metal enhanced region in the northern rim of the Loop.  

\section{Observations}
We summarized the 5 observations in table \ref{tab:info}. Their FOV are shown in figure \ref{fig:region} left panel with the white solid squares. We also show the FOV of the other rim observations with the white dotted squares: four pointings of the northeastern rim (NE1-NE4; \cite{Katsuda08NE}) and a SE rim observation (P27; \cite{Tsunemi09}). We intended to expand our observation westward from NE1-NE4 along the rim. We note that the FOV of P21 is next to that of NE4.

All the data were analyzed with version 6.5.1 of the HEAsoft tools. For the reduction of these data, we used the version 9 of the Suzaku Software. The calibration database (CALDB) used was the one updated in September 2008. We used the revision 2.2 of the cleaned event data and combined the 3$\times$3 and 5$\times$5 event files. All the data were taken by using the spaced row charge injection (SCI) method \citep{Prigozhin08} which reduces the effect of radiation damage of the XIS and recovers the energy resolution, for example, from 205$\pm6$eV to 157$\pm4$eV at the He-like Fe K line. In order to exclude the background flare events, we obtained the good time intervals (GTIs) by including only times at which the count rates are within $\pm2\sigma$ of the mean count rates.

The Cygnus Loop is a large diffuse source and our FOV are almost filled with the SNR's emission. We also have no background data from the neighborhood of the Cygnus Loop. We therefore applied the Lockman Hole data for the background subtraction. We reviewed the effect of the Galactic Ridge X-ray Emission (GRXE). The flux of the GRXE at $l = 62^\circ$, $|b| < 0^\circ.4$ is $6\times10^{-12}$\,erg$\cdot$cm$^{-2}$s$^{-1}$deg$^{-2}$ (0.7-2.0\,keV) \citep{Sugizaki01}. Although the Cygnus Loop ($l = 74^\circ$, $b = -8^\circ.6$) is located outside of the FOV of \citet{Sugizaki01}, this value gives us an upper limit of the GRXE at the Cygnus Loop. Meanwhile, the total count rate of the Cygnus Loop is 1659 counts/s (0.8-2.01\,keV) \citep{Aschenbach99}, the flux is estimated to be $1.68\times10^{-10}$\,erg$\cdot$cm$^{-2}$s$^{-1}$deg$^{-2}$, assuming that the effective area of the \textit{ROSAT} HRI is 80\,cm$^2$ and that the Cygnus Loop is a circle of $3^\circ.0$ in diameter. This value is consistent with the results from our FOV. Therefore, we concluded that the effect of the GRXE on the Cygnus Loop is vanishingly small. The solar wind charge exchange (SWCX) is also considered to be one of the correlates of the soft X-ray background below 1\,keV  \citep{Fujimoto07}. However, in terms of the Cygnus Loop, we consider that the SWCX is negligible because of the prominent surface brightness of the Loop. Thus, for the background subtraction, the Lockman Hole data obtained in May 2008 were applied. These observation dates were close to those of the Cygnus Loop observations and we confirmed that they have no background flares below 1\,keV. Since there were no photons above 3.0\,keV after the background subtraction, the energy ranges of 0.2-3.0\,keV and 0.4-3.0\,keV were used for XIS1 (back-illuminated CCD; BI CCD) and XIS0,3 (front-illuminated CCD; FI CCD), respectively \citep{Koyama07}. 

\begin{table}
 \begin{center}
 \caption{Summary of the 5 observations}\label{tab:info}
  \begin{tabular}{lcccc}
\hline\hline
Obs. ID & Obs. Date& Coordinate (J2000) (RA, DEC) & Position Angle & Effective Exposure\\
\hline\hline
503057010  (P21) & 2008-06-02 &  \timeform{20h52m43.8s}, \timeform{32D26'19.0''} & 61$^\circ$.9 & 16\,ksec\\
\hline
503058010  (P22) & 2008-06-03 &  \timeform{20h51m17.2s}, \timeform{32D25'24.6''} & 61$^\circ$.4 & 5.9\,ksec\\
\hline
503059010  (P23) & 2008-06-03 &  \timeform{20h49m50.6s}, \timeform{32D21'50.8''} & 61$^\circ$.9 & 19\,ksec\\
\hline
503060010  (P24) & 2008-06-04 &  \timeform{20h48m28.2s}, \timeform{32D17'44.5''} & 61$^\circ$.4 & 10\,ksec\\
\hline
503061010  (P25) & 2008-06-04 &  \timeform{20h47m22.7s}, \timeform{32D10'22.8''} & 60$^\circ$.9 & 21\,ksec\\
\hline
  \end{tabular}
\end{center}
\end{table}

\section{Spectral Analysis}
\begin{figure}
  \begin{center}
    \FigureFile(76mm,0mm){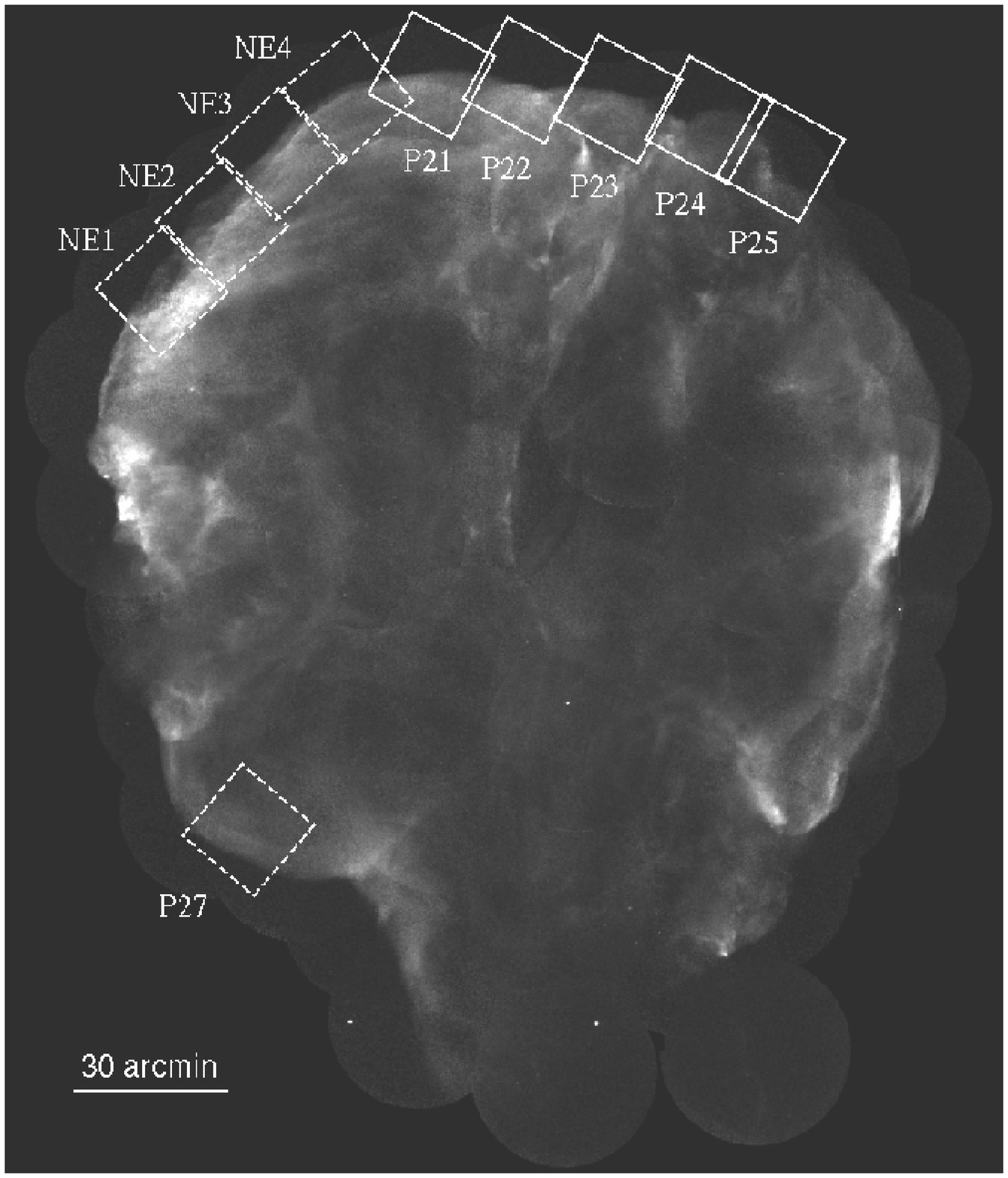}
    \FigureFile(68mm,0mm){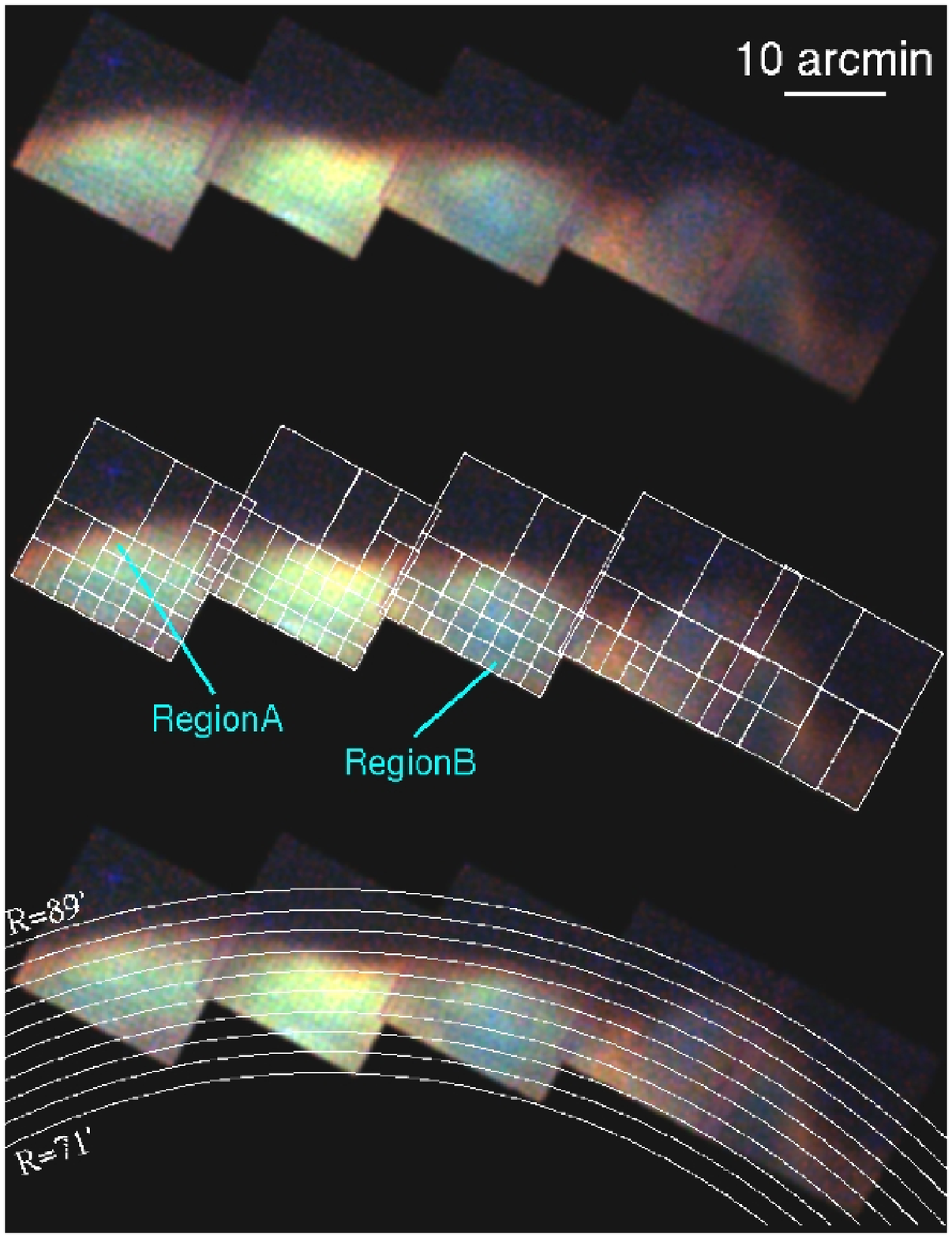}
  \end{center}
  \caption{\textit{Left}: \textit{ROSAT} HRI image of the entire Cygnus Loop. The \textit{Suzaku} FOV are shown with white rectangles. The dotted and solid lines represent the past and our observations, respectively. \textit{Right top}: Three-color X-ray image for P21-P25 using XIS data. Red, green and blue correspond to the energy ranges of 0.3-0.5\,keV, 0.5-1.0\,keV and 1.0-3.0\,keV, respectively. \textit{Right middle}: Same as the right top panel, but for overlaid with the spectral extraction regions with white rectangles. \textit{Right bottom}: Same as the right middle panel, but for the different spectral extraction regions.}\label{fig:region}
\end{figure}

Figure \ref{fig:region} right shows the three-color X-ray image for P21-P25 using \textit{Suzaku} XIS data after correcting for exposure and vignetting effects. 
Red, green and blue correspond to the energy ranges of 0.3-0.5\,keV, 0.5-1.0\,keV and 1.0-3.0\,keV, respectively. 
To investigate the plasma structure of the northern rim of the Loop, we divided our FOV into several box regions as shown in figure \ref{fig:region} right middle. 
In order to equalize the statistics, we initially divided all images of XIS1 into two parts and if each divided region has more than 10,000 photons, it was once again divided. 
In this way, we obtained 115 box regions. 
Each region contains 5,000-10,000 photons for XIS1. 
The side length of each box ranges from 2${^\prime}$.3 to 9${^\prime}$.0. 
Therefore, box sizes are not smaller than the angular resolution capability of the \textit{Suzaku} XIS. 
We grouped 115 spectra into bins with a minimum of 20 counts so that $\chi^2$ statistics is appropriate. 
In order to generate a response matrix file (RMF) and an ancillary response file (ARF), we employed \textbf{xisrmfgen} and \textbf{xissimarfgen} \citep{Ishisaki07} for our data. 
We generated ARFs for ``extended sources''.

 We first applied an absorbed single-$kT_e$ component of non-equilibrium ionization (NEI) plasma model for all the spectra. 
 We employed \textbf{TBabs} (Tuebingen-Boulder ISM absorption model; \cite{Wilms00}) and \textbf{VNEI} (NEI ver.2.0; \cite{Borkowski01}) in XSPEC version 12.4.0 \citep{Arnaud96}. 
In this model, the abundances of C, N, O, Ne, Mg, Si and Fe were free while we set the relative abundances of S to the solar value \citep{Anders89} equal to that of Si, Ni equal to Fe. Other elements were fixed to their solar values. 
Other parameters were all free such as the electron temperature $kT_e$, the ionization timescale $\tau$ (a product of the electron density and the elapsed time after the shock heating), and the emission measure (EM $= \int n_e n_{\rm H} dl$, where $n_e$ and $n_{\rm H}$ are the number densities of hydrogen and electron, $dl$ is the plasma depth). 
We also set the column density $\rm\textit{N}_H$ free. 
The spectra are reasonably well fitted by the single-$kT_e$ VNEI model for almost all regions. 
The values of the reduced $\chi^2$ show around 1.5 and the degrees of freedom (dof) are 300-400. 
We also fitted all the spectra with the two-$kT_e$ VNEI model in which the values of the abundances and $\tau$ were tied in two components. 
This model was applied for the NE rim observations in \citet{Miyata07} and \citet{Katsuda08NE}. 
However, this model does not improve the values of $\chi^2$ for almost all regions and the best-fit parameters are consistent with those obtained from the single-$kT_e$ VNEI model. 
Therefore, we concentrate on the results of the single-$kT_e$ VNEI model in what follows. 

\begin{figure}
  \begin{center}
    \FigureFile(70mm,0mm){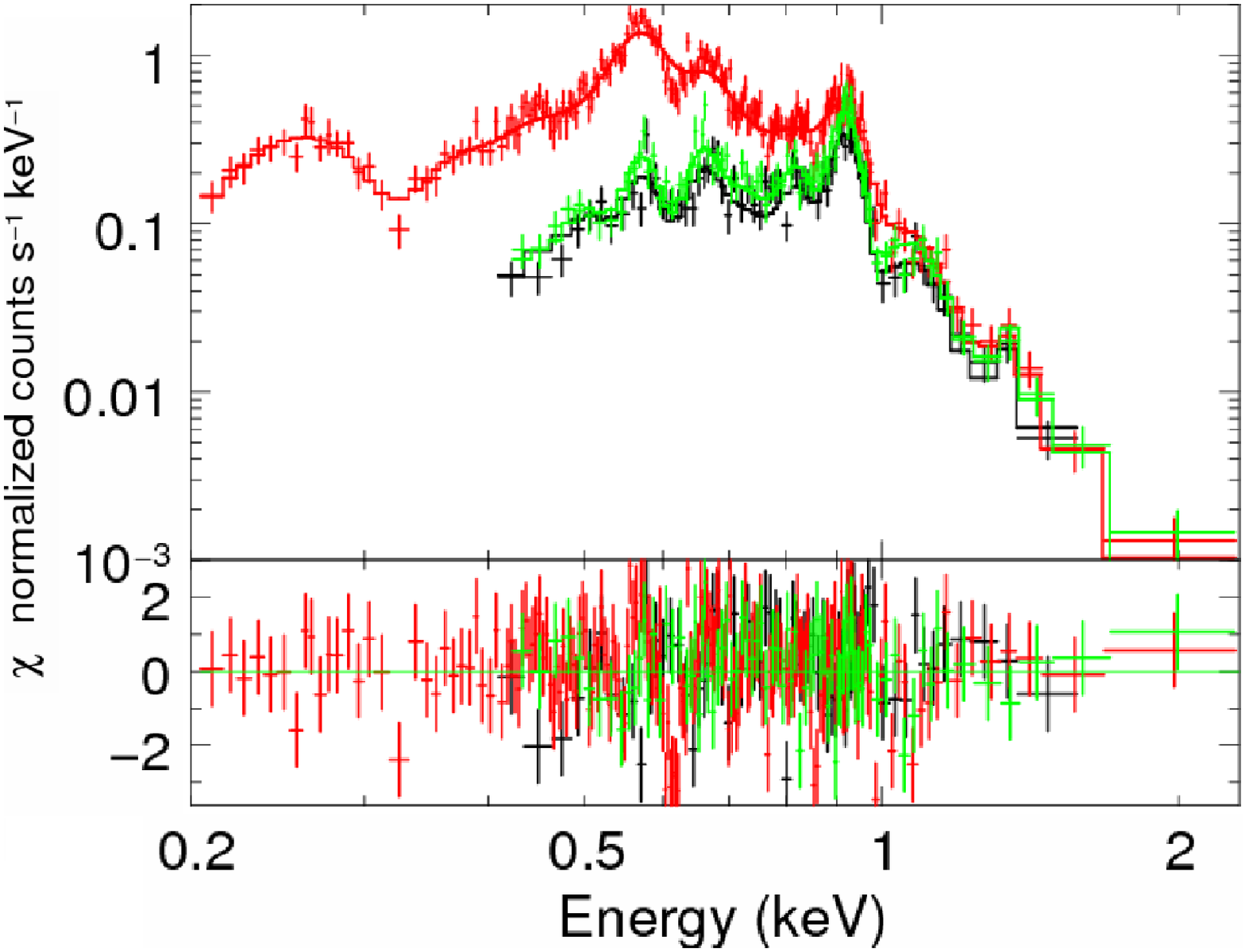}
    \FigureFile(72mm,0mm){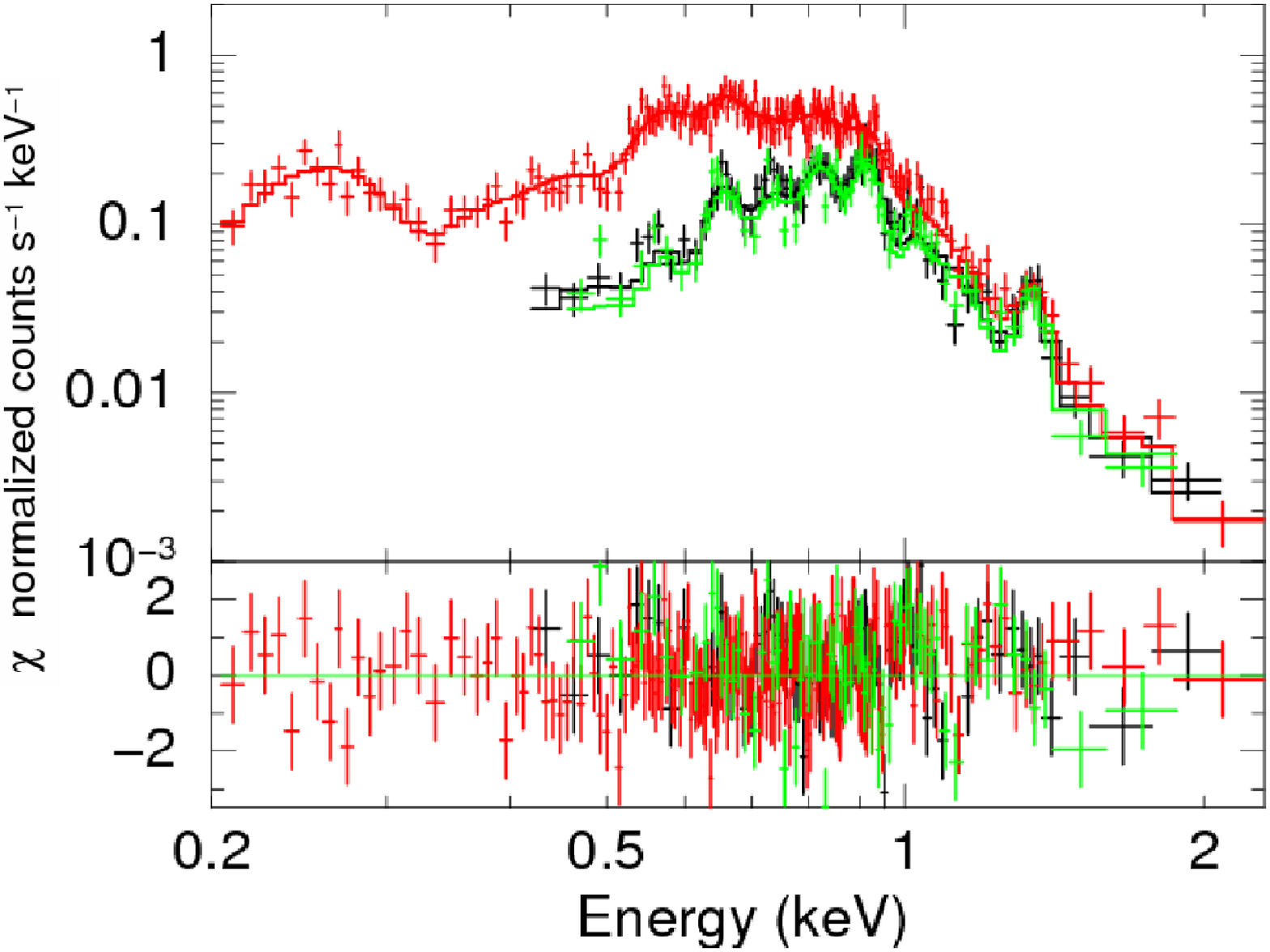}
  \end{center}
  \caption{Example spectra from the regions where N, O and Ne are abundant (region-A: left) and depleted (region-B: right), respectively. The best-fit curves are shown with solid line.  The residuals are shown in the lower panels. Black, red, green correspond to the XIS 0, 1, 3, respectively.}\label{fig:spec1}
\end{figure}

The example spectra and the best-fit curves are shown in figure \ref{fig:spec1}. 
The best-fit parameters are shown in table \ref{tab:spec1}.  
We also show the spectral extracted regions in figure \ref{fig:region} right middle. 
These two spectra are taken from the regions where N, O and Ne are relatively abundant (region-A) and depleted (region-B). 
From figure \ref{fig:spec1}, we can clearly see the difference in the feature of each spectrum: the spectra obtained from region-B is smoother than that from region-A and the line emissions of some elements such as N, O and Ne are prominent in region-A. 
The best-fit parameters in table \ref{tab:spec1} support these differences statistically.
The abundances of N, O and Ne are significantly higher in region-A than those in region-B.
Figure \ref{fig:paramap} shows the maps of the best-fit parameters obtained from all 115 spectra. 
The images are smoothed by Gaussian kernel of $\sigma$=1.5${^\prime}$. 
The color code scale is normalized by the maximum values. 
From figure \ref{fig:paramap}, the values of $kT_e$ distribute around 0.3\,keV and those in the inner region of P23 are higher than the other parts of our FOV. 
The values of log($\tau$) are higher in the west than those in the east and spread between 10.5-11.5.
Some elements such as N, O and Ne are particularly abundant in the outer edge of the rim in P21-P22 while they seem to be depleted inward.
These results suggest that the abundances of these elements have radial dependencies. 
Then, we determined the spectral extraction regions in different way to investigate the plasma structure from the inner side to the outer edge of the rim. 

\begin{figure}
  \begin{center}
    \FigureFile(80mm,0mm){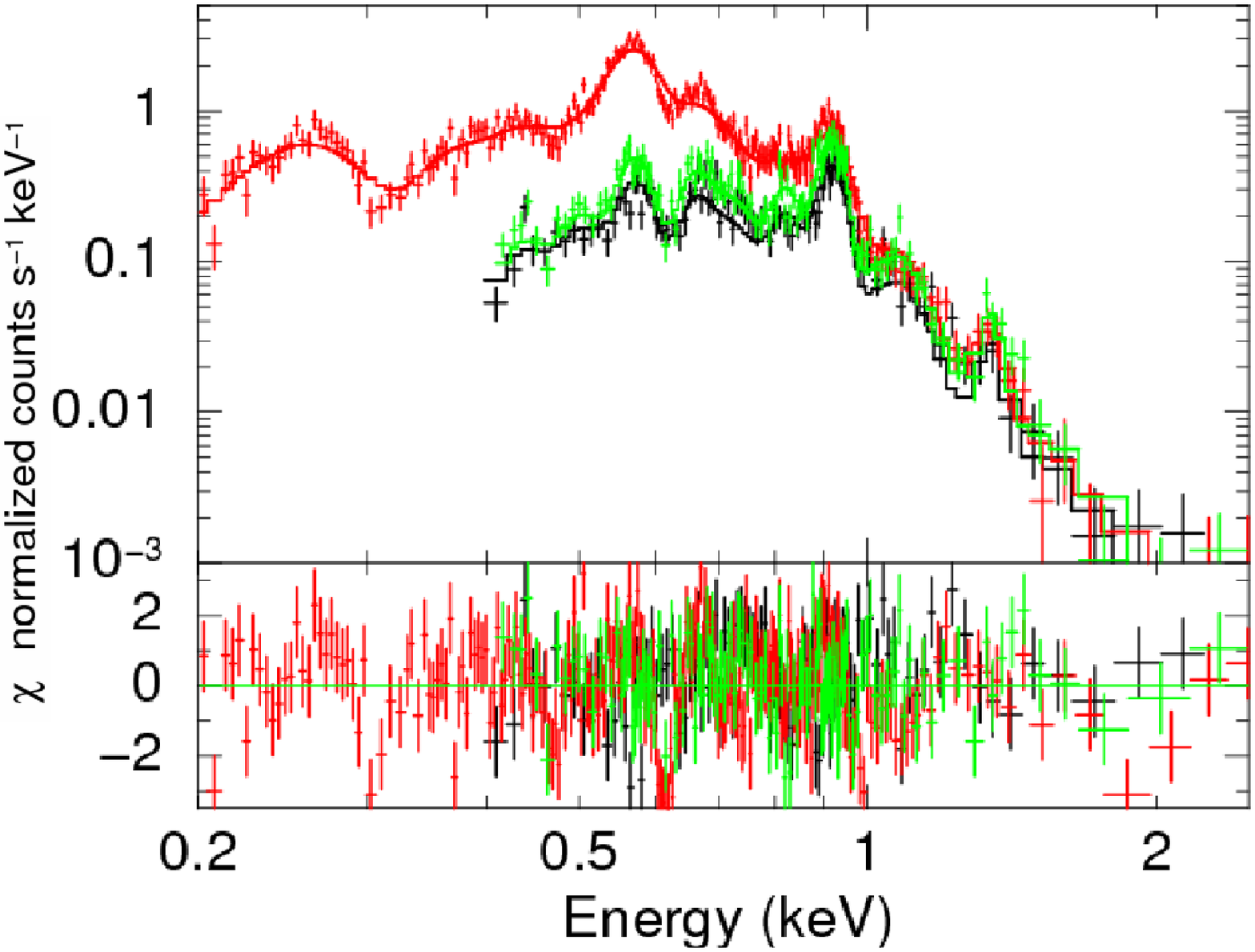}
    \FigureFile(80mm,0mm){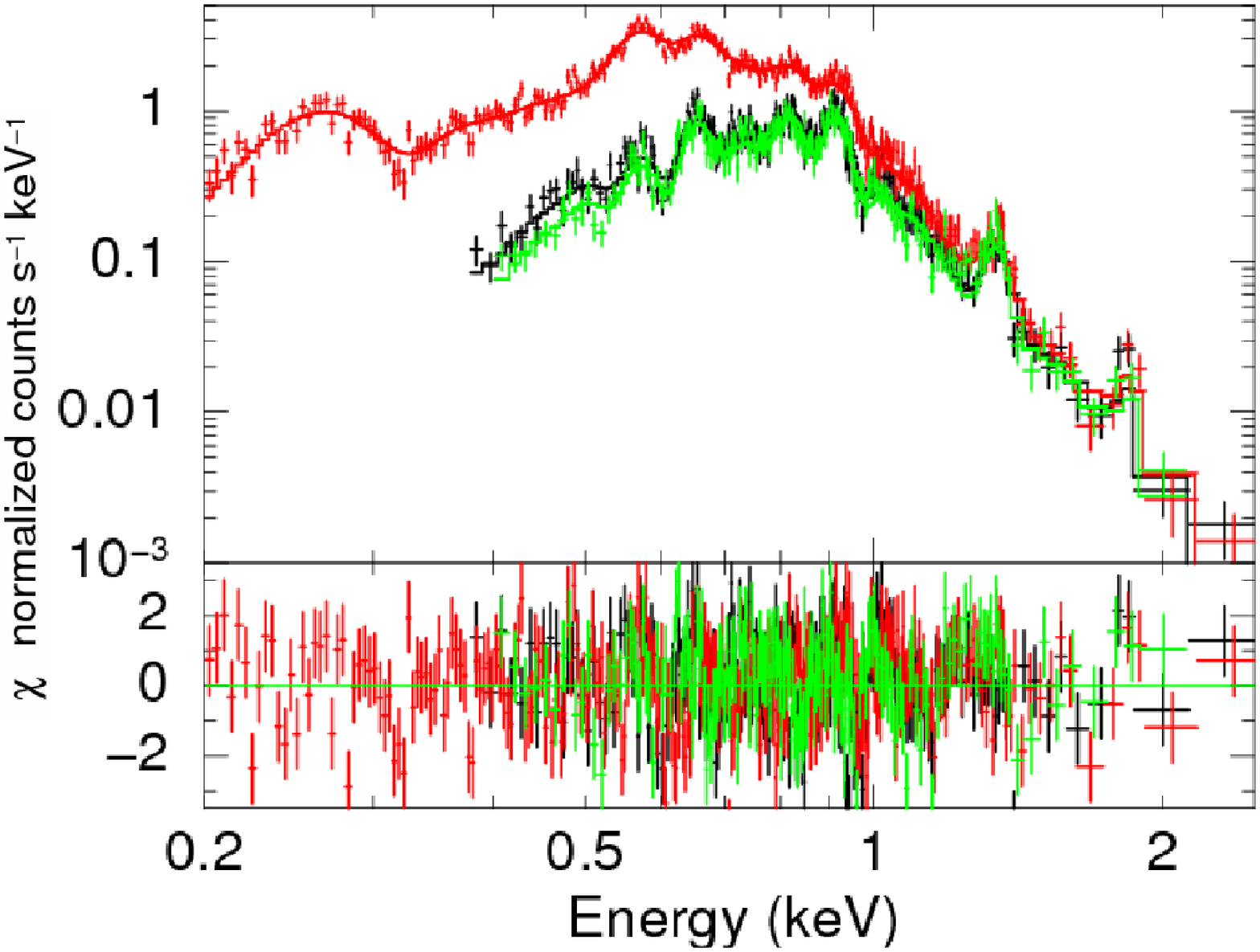}
  \end{center}
  \caption{Example spectra extracted from the outer edge of the rim ($R$=88${^\prime}$) and the inner region ($R$=76${^\prime}$). The best-fit curves are shown with solid line.  The residuals are shown in the lower panels. Black, red, green correspond to the XIS 0, 1, 3, respectively.}\label{fig:spec2}
\end{figure}

As shown in figure \ref{fig:region} right bottom, we divided our FOV into several annular sectors.
We set the center of the annular regions to the ``geometric center'' estimated by \citet{Levenson98}. 
It is located on $\alpha = $\timeform{20h51m21s}, $\delta = $\timeform{31D01'37''}, determined by fitting the \textit{ROSAT} HRI map of the Cygnus Loop by the model circle.
We set the annular width to 2${^\prime}$ which is restricted by the angular resolution capability of the \textit{Suzaku} XIS.
The distance from the geometric center ranges from 72${^\prime}$ to 88${^\prime}$.
In this way, we have totally 36 annular sectors from 5 observations.
Figure \ref{fig:spec2} shows two examples of the spectra from P21. 
One is extracted from the outer edge of the rim ($R$=88${^\prime}$, where $R$ represents the angular distance from the center) and the other, extracted from the inner region ($R$=76${^\prime}$). 
As is the case in figure \ref{fig:spec1}, the emission lines of N, O, and Ne appear more in the outer region ($R$=88${^\prime}$) than those in the inner one ($R$=76${^\prime}$). 
The best-fit curves and the best-fit parameters with the single-$kT_e$ VNEI model are shown in figure \ref{fig:spec2} and table \ref{tab:spec2}, respectively.
From table \ref{tab:spec2}, we found that the abundances of some elements such as N, O, Ne, Mg and Fe are significantly higher at $R$=88${^\prime}$ than those at $R$=76${^\prime}$.
Figure \ref{fig:funcr} shows the radial plot of the best-fit parameters, $kT_e$, log($\tau$), and abundances of various elements as a function of $R$.
We found that the abundances of N, O, and Ne increase toward the outer edge of the rim only in P21-P22. 
This result is consistent with that of figure \ref{fig:paramap}. 

\begin{table}
  \begin{center}
 \caption{Spectral fit parameters}\label{tab:spec1}
    \begin{tabular}{lcc}
       \hline 
      \hline
 Parameter & region-A (VNEI) & region-B (VNEI)\\
      \hline
      N$\rm _H$ [10$^{20}$cm$^{-2}$]  & 1.0 $\pm$ 0.1 & 1.0 $\pm$ 0.1 \\
       $kT_e$ [keV]  & 0.32 $^{+0.04}_{-0.07}$ & 0.43 $^{+0.04}_{-0.05}$ \\
       C  & 0.81 $^{+2.79}_{-0.43}$ & 0.18 $^{+0.16}_{-0.13}$\\
       N  & 0.61 $^{+0.72}_{-0.45}$ & $<$ 0.06\\
       O  & 0.45 $^{+1.03}_{-0.26}$ & 0.10 $^{+0.03}_{-0.02}$\\
       Ne  & 1.04 $^{+2.03}_{-0.64}$ & 0.24 $^{+0.06}_{-0.04}$\\
       Mg  & 0.56 $\pm$ 0.18 & 0.25 $^{+0.08}_{-0.06}$\\
       Si (=S)  & $<$ 0.17 & 0.18 $^{+0.12}_{-0.10}$\\
       Fe (=Ni)  & 0.58 $^{+0.83}_{-0.39}$ & 0.21 $^{+0.05}_{-0.03}$\\
       log($\tau$)  & 10.52 $^{+0.37}_{-0.07}$ & 10.79  $^{+0.37}_{-0.14}$\\
       EM [10$^{20}$cm$^{-5}$]  & 0.24 $\pm$ 0.01 & 0.23 $\pm$ 0.09\\
$\chi ^2$/dof  & 672/437 & 398/322\\
      \hline
    \end{tabular}
 \end{center}
\end{table}

\section{Discussion and Conclusion}
\subsection{Abundance depletion at the rim of the Cygnus Loop}

\citet{Miyata94} observed the NE rim of the Loop (overlapping with the NE2 region) with \textit{ASCA} and found the deficient metal abundances there (typically $\sim$0.1 times solar). 
\citet{Miyata07} confirmed it with the \textit{Suzaku} satellite.
The abundance depletion is also seen at the other side of the Loop \citep{Leahy04}.
\citet{Leahy04} observed SW rim of the Loop with \textit{Chandra} and showed that the abundance of the O-group (C, N and O) is depleted about twice as that of the Ne-group (Ne, Mg, Al, Si, S and Ar) and the Fe-group (Ca, Fe and Ni). 
He showed the O-group abundance to be 0.22 times the solar value.
Likewise, several other X-ray studies reported such low metal abundances (\cite{Miyata98}; \cite{Miyata99}; \cite{Tsunemi07}; \cite{Katsuda08NE}; \cite{Katsuda08Chan}; \cite{Tsunemi09}).
It seems to be a common result at the rim of the Cygnus Loop.
Our result confirmed this trend at the northern rim of the Loop, except the outermost part of P21-P22 (\textit{e.g.} see figure \ref{fig:paramap}).

\citet{Cartledge04} observed the O and H absorption line and measured the interstellar O along 36 sight lines.
The results indicate that the O/H ratio is homogeneous within 800pc of the sun and that the O abundance is $\sim$0.4 times the solar value.
Since the Cygnus Loop is located at 540pc of the sun, the O abundance is expected to be that of the interstellar medium (ISM).
The past observations do not show such values and it is not clear why the O abundance and the other elemental abundances are all deficient at the rim of the Loop.
 
\citet{Raymond03} studied the effect of the resonance-scattering for the O VI emission from the FUV observations in the NE rim of the Cygnus Loop.
\citet{Miyata08} proposed that the effect of the resonance-line-scattering optical depth lowers the apparent abundances of some elements on the basis of the NE2 observation. 
They calculated the optical depths in their FOV for some K emission lines (C VI K$\alpha$, N VI K$\alpha$, O VII K$\alpha$, O VIII Ly$\alpha$, and Ne IX K$\alpha$) and concluded that the optical depth effects play a significant role in the X-ray emission lines, particularly those of O. 
They calculated that the O abundance is underestimated by a factor of 20-$40\%$ in the NE rim where the O emission is the highest in the Cygnus Loop. 
However, it is not sufficient to account for the abundance depletion in the NE rim.
The resonance-line-scattering effect depends linearly on the column density, the product of the density and the plasma depth.
We here assume that the Cygnus Loop is spherical symmetric and the filling factor in the shell region is unity. 
Then, the plasma depths in the NE rim and our FOV are the same with each other, since these regions are located at about the same radii. 
Therefore, the resonance-line-scattering effect depends only on the density which in this case squarerootly depends on the EM or the surface brightness. 
Since the surface brightness in our FOV is lower than that in the NE rim, the resonance-line-scattering effect should be even less significant in our FOV than that in the NE rim. 
Therefore, we can conclude that the effect of the resonance-line-scattering is not sufficient enough to account for the abundance depletion observed in our FOV.

The other possibility is the effect of the circumstellar grain.
In general, the grain condensation depletes the abundance of the heavy elements such as C, O, Mg, Si and Fe \citep{Savage96}.
The cavity wall is considered to be an ISM material pushed into a shell by the stellar wind and radiation while the progenitor was on the main sequence. 
There is a possibility that the ISM grains were destroyed by shocks as the shell expanded \citep{Vancura94} and that the condition of the abundance of grain influences the values of the metal abundances.
However, this model does not explain the depletion of rare gasses such as Ne and it remains unclear why the metal abundances show low values almost everywhere in the cavity wall.

\subsection{Abundance-enhanced region at the outermost edge of the Cygnus Loop}

On the other hand, some recent observations revealed that the limited regions show high abundances (\cite{Katsuda08NE}; \cite{Tsunemi09}).
\citet{Katsuda08NE} analyzed the \textit{Suzaku} data of the NE rim (NE1-NE4) and found that the abundances of the heavy elements have relatively high values (C$\sim$0.7, N$\sim$0.7, O$\sim$0.4, Ne$\sim$0.6, Mg$\sim$0.3 and Fe$\sim$0.3) only at the outermost edge in NE3-NE4, while the other regions show relatively depleted abundances. 
They also showed the abundance ratios of Mg/O and Fe/O are lower in the abundance-enhanced region than those in the other regions.
\citet{Katsuda08Chan} made an additional observation with \textit{Chandra} and confirmed these results.
The abundance-enhanced region is about 30$^\prime\times$3$^\prime$ with relatively weak surface brightness.
\citet{Tsunemi09} observed the SE rim (P27) with \textit{Suzaku} and showed that the similar abundance-enhanced regions are seen in their FOV (C$\sim$0.6, N$\sim$0.9, O$\sim$0.4, Ne$\sim$0.7, Mg$\sim$0.5 and Fe$\sim$0.5). 
The width of the abundance-enhanced region is $\sim$3$^\prime$.
It is striking that these regions are both located at the outermost edges of the rims, while the inner regions usually show the metal deficient abundances like the other rim observations.

\begin{figure}
  \begin{center}
    \FigureFile(70mm,0mm){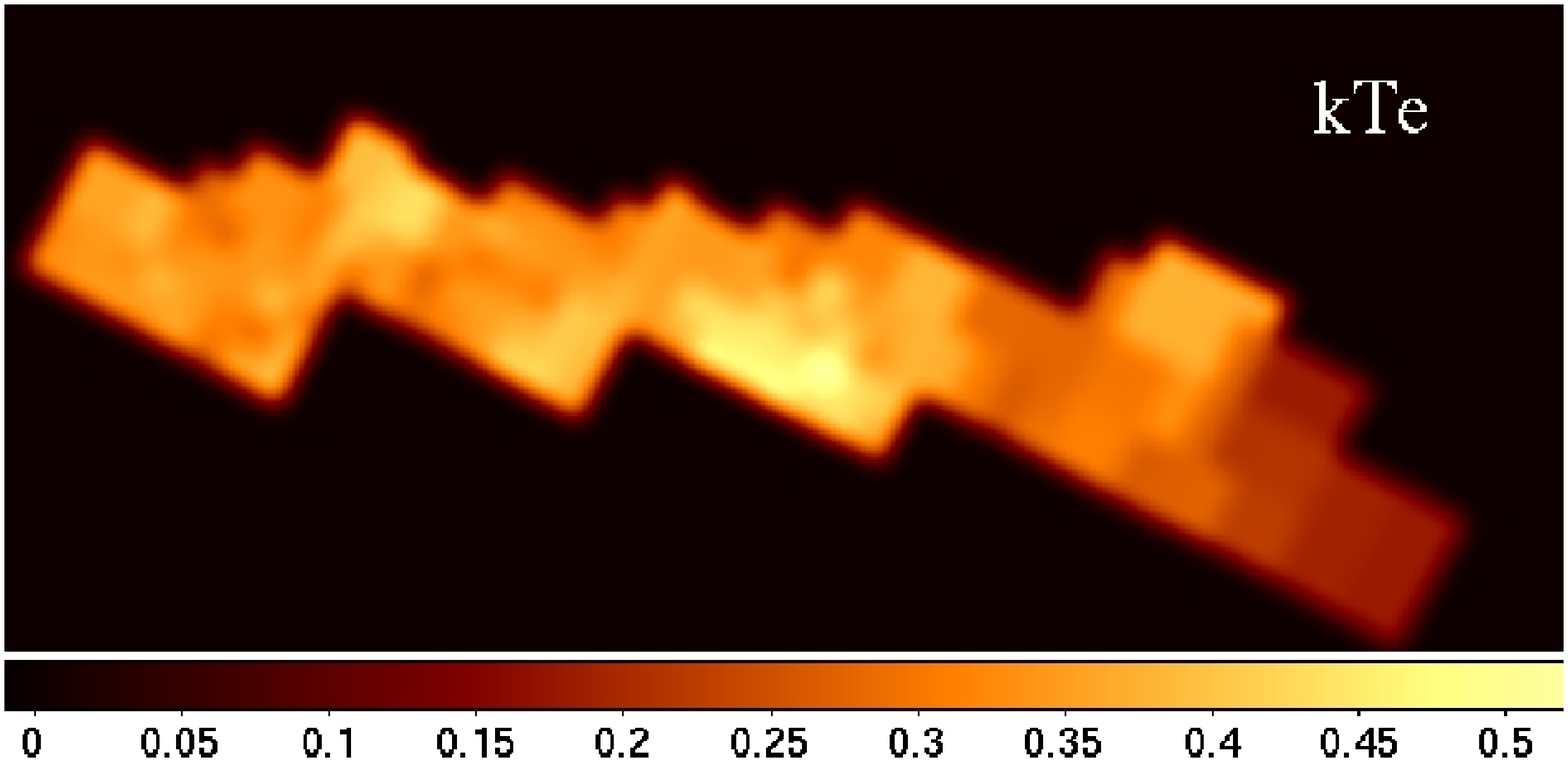}
    \FigureFile(70mm,0mm){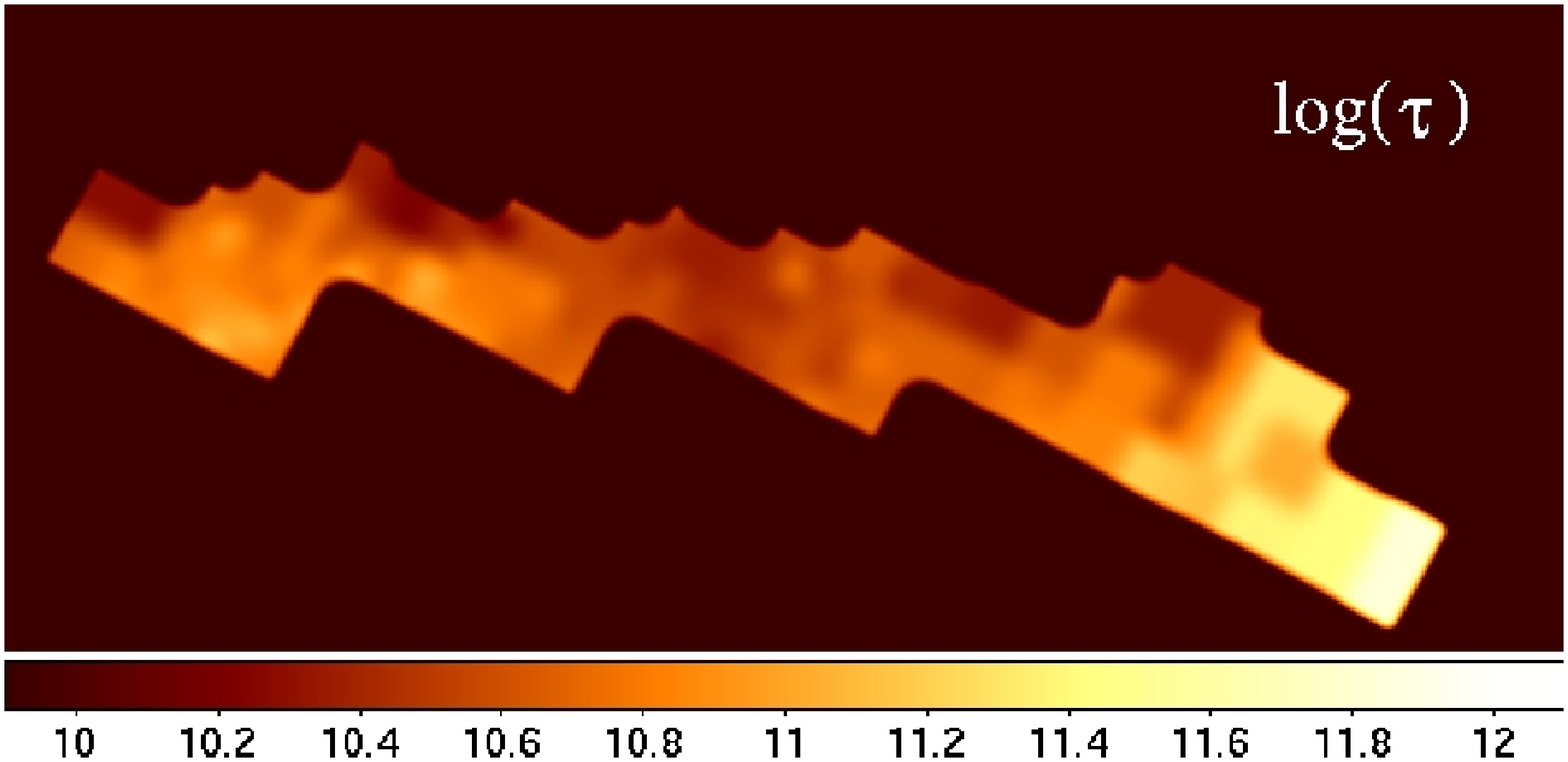}
    \FigureFile(70mm,0mm){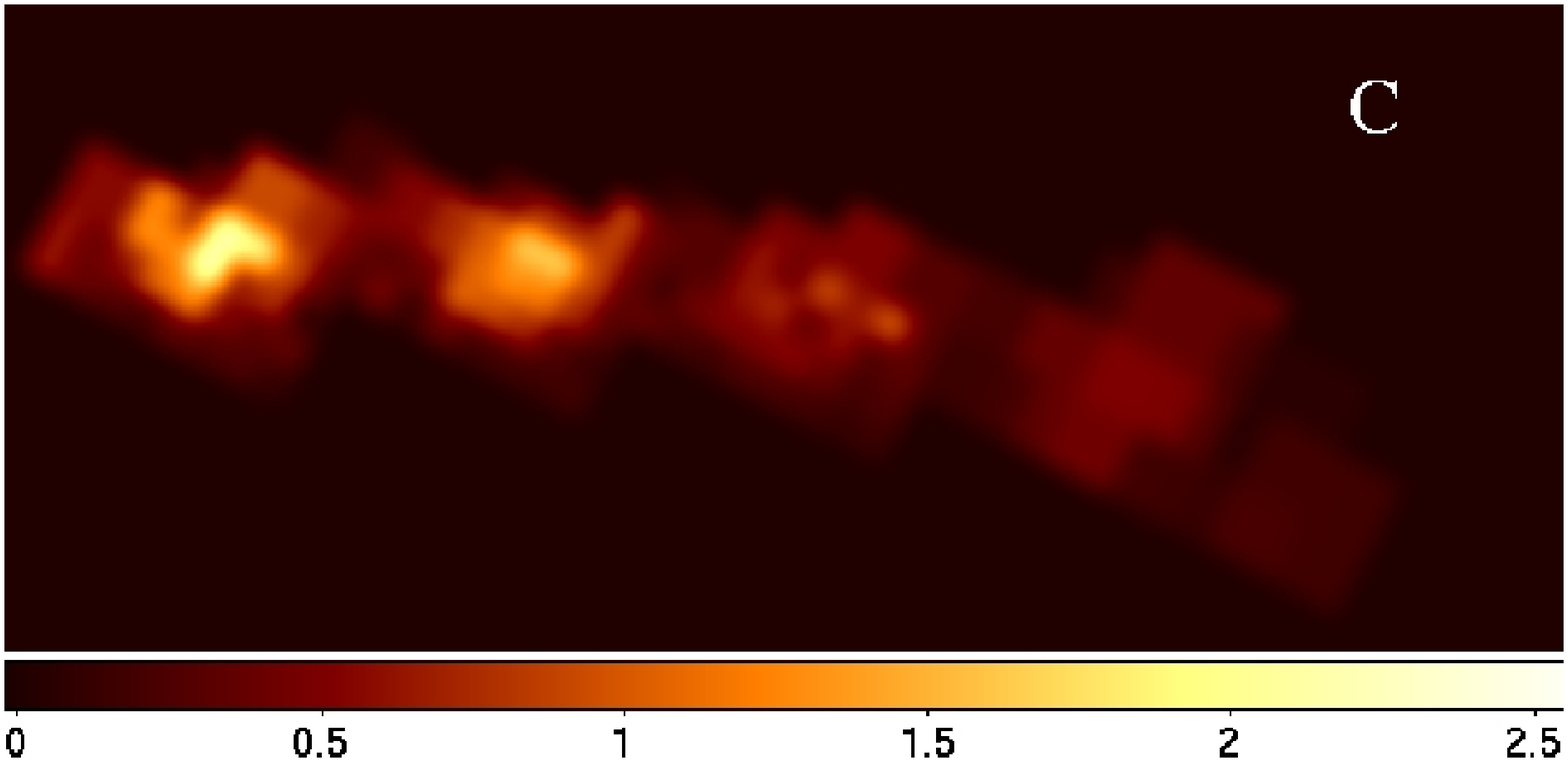}
    \FigureFile(70mm,0mm){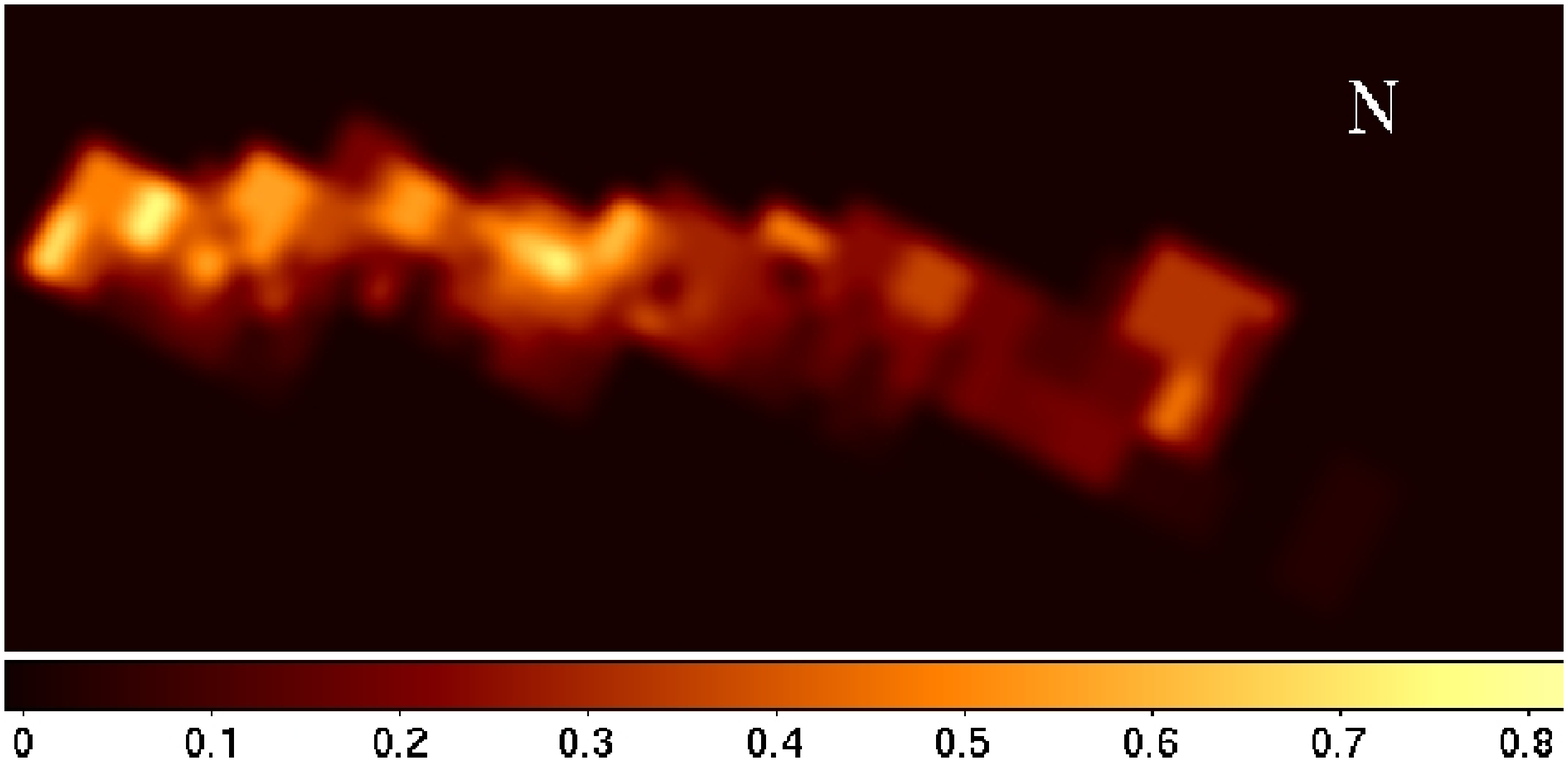}
    \FigureFile(70mm,0mm){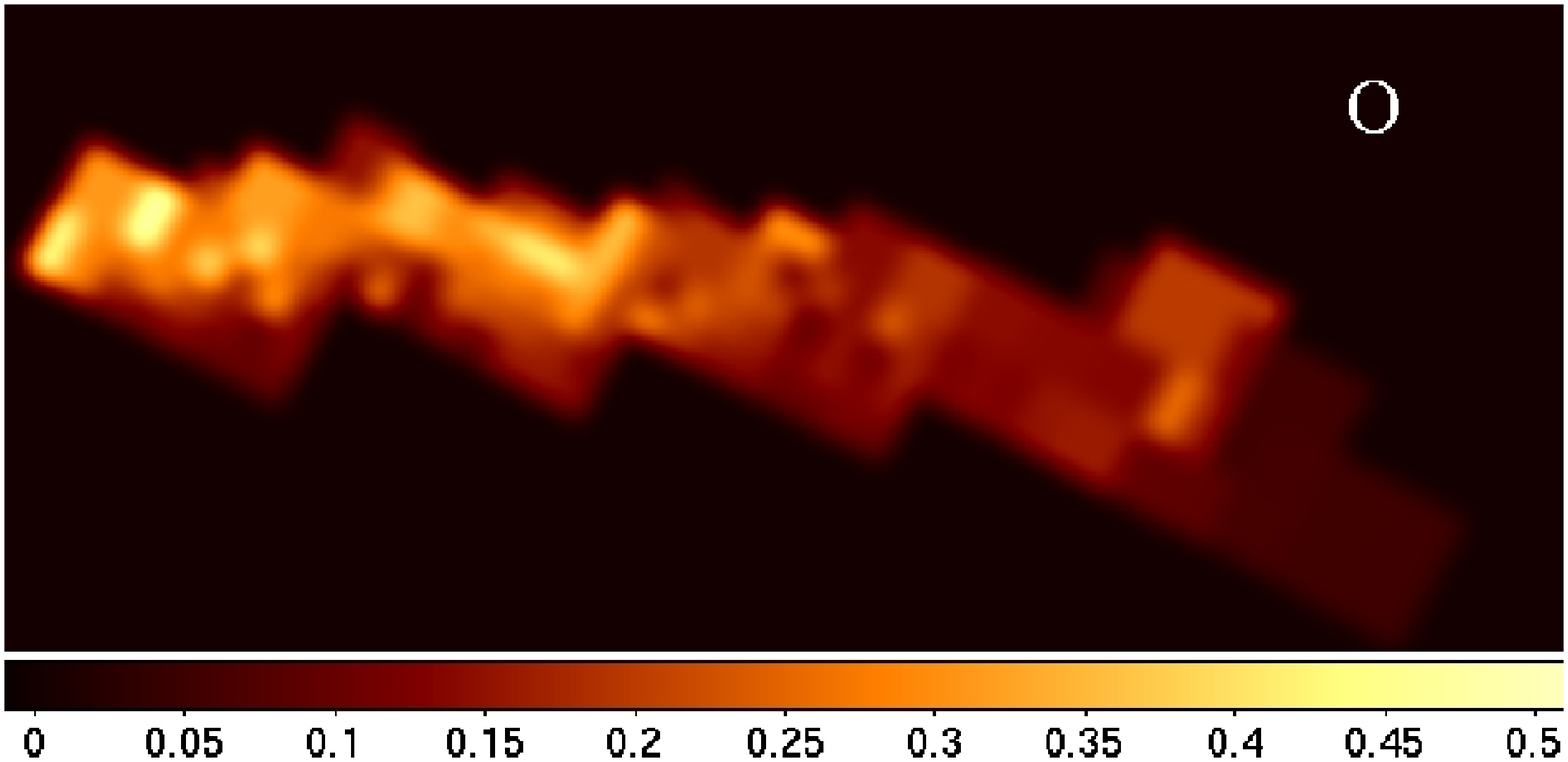}
    \FigureFile(70mm,0mm){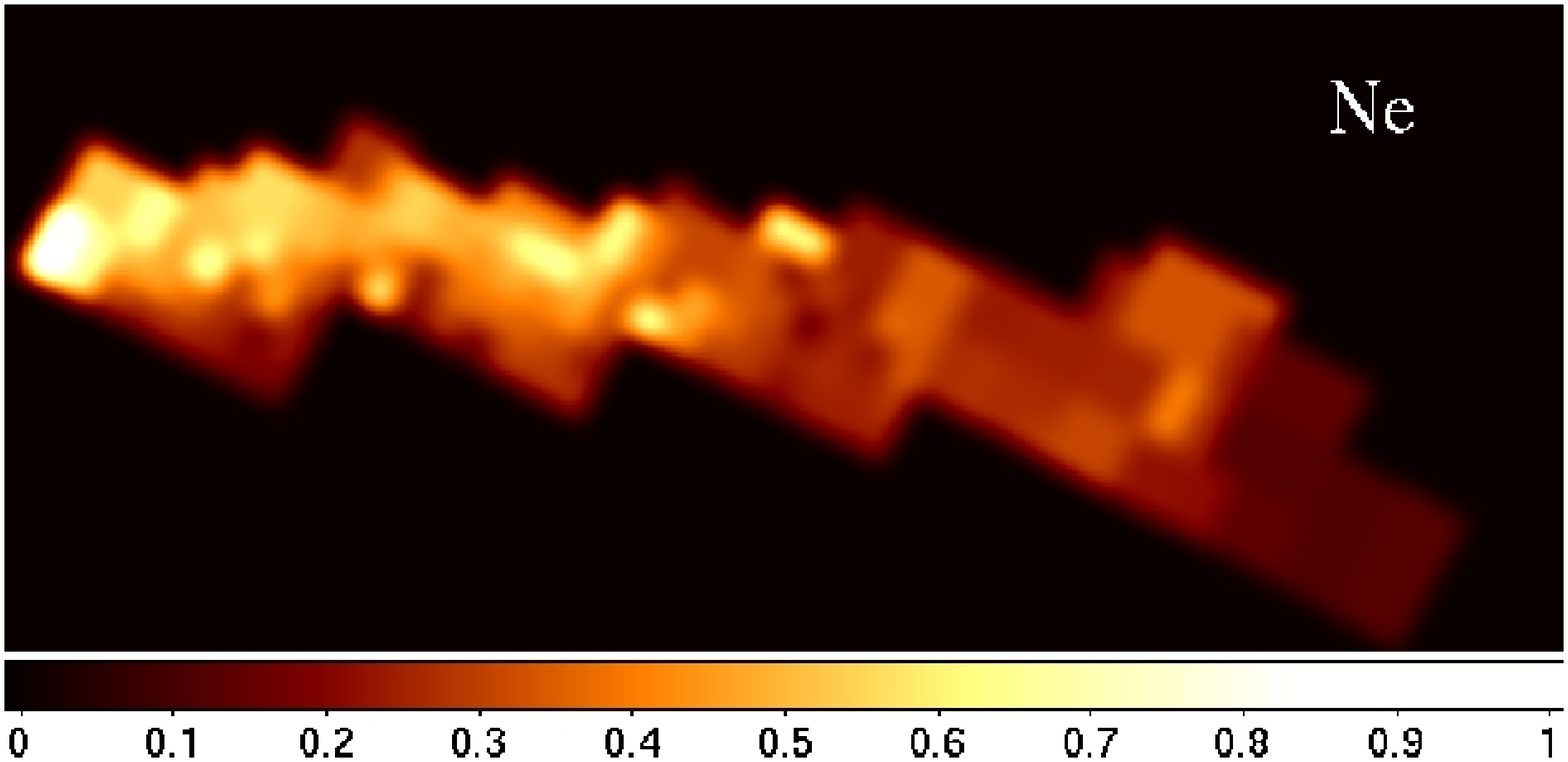}
    \FigureFile(70mm,0mm){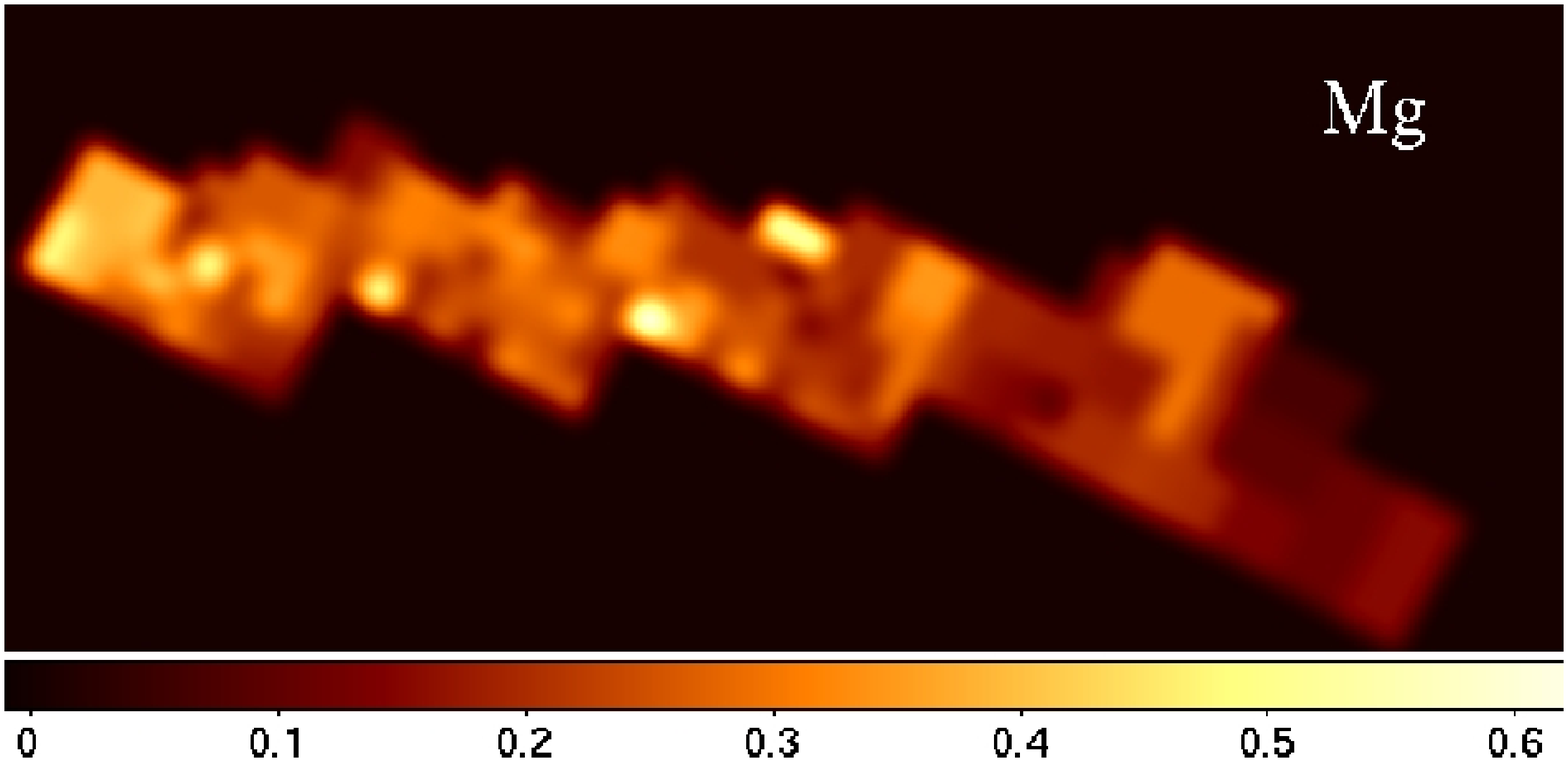}
    \FigureFile(70mm,0mm){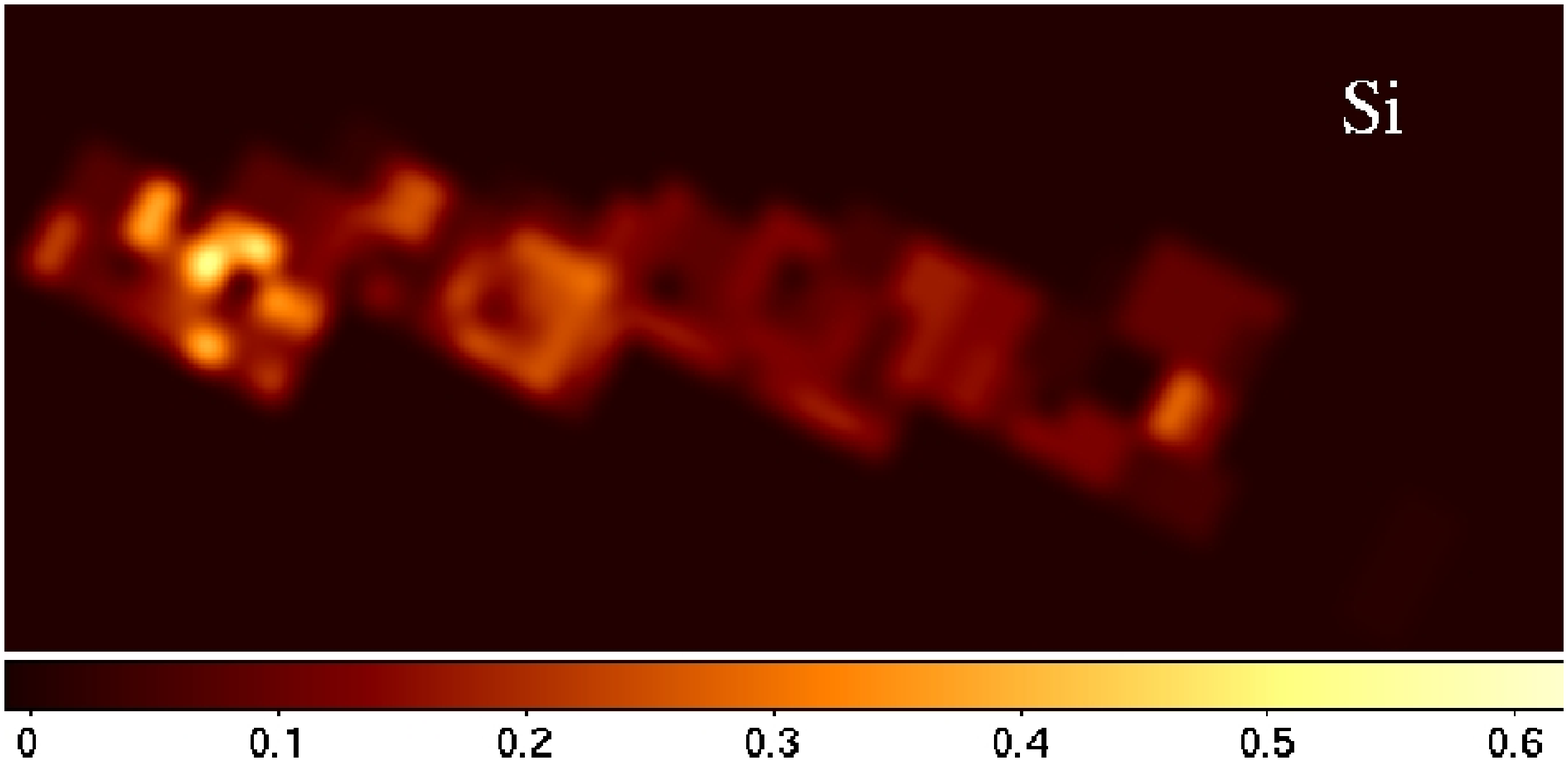}
    \FigureFile(70mm,0mm){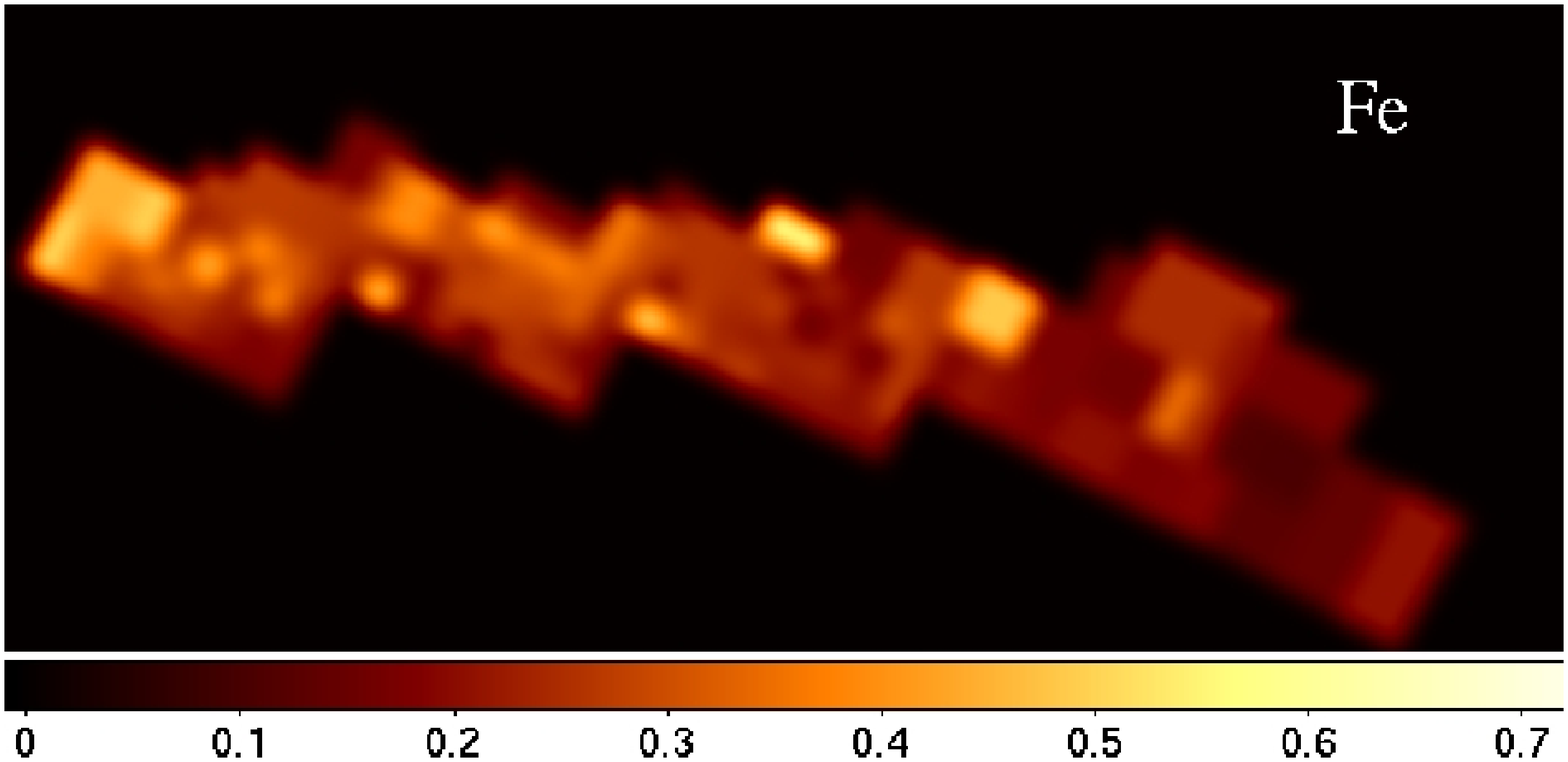}
  \end{center}
  \caption{Maps of the best-fit values for various parameters. The value of $kT_e$ is in units of keV.}\label{fig:paramap}
\end{figure}

\begin{figure}
  \begin{center}
    \FigureFile(60mm,0mm){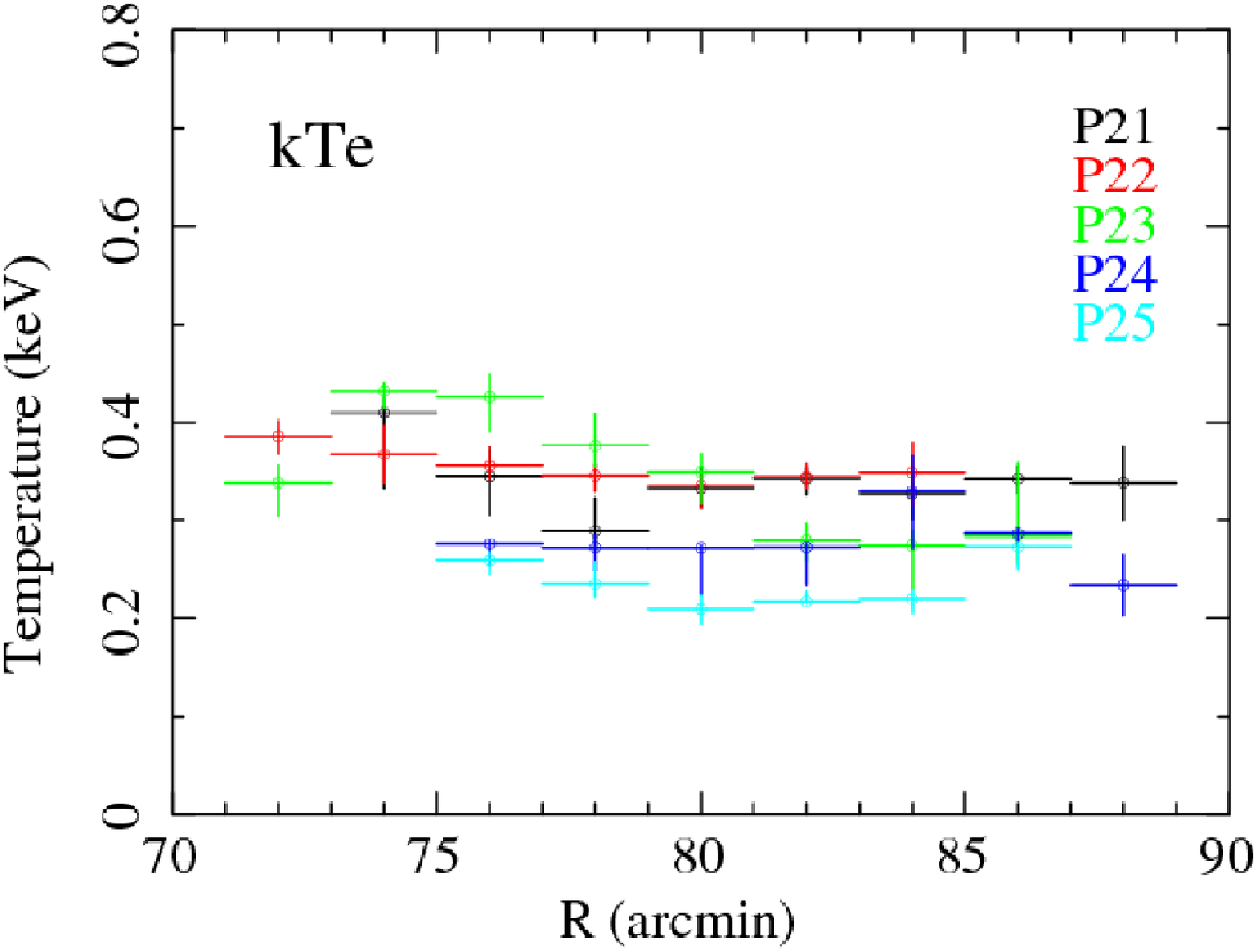}
    \FigureFile(60mm,0mm){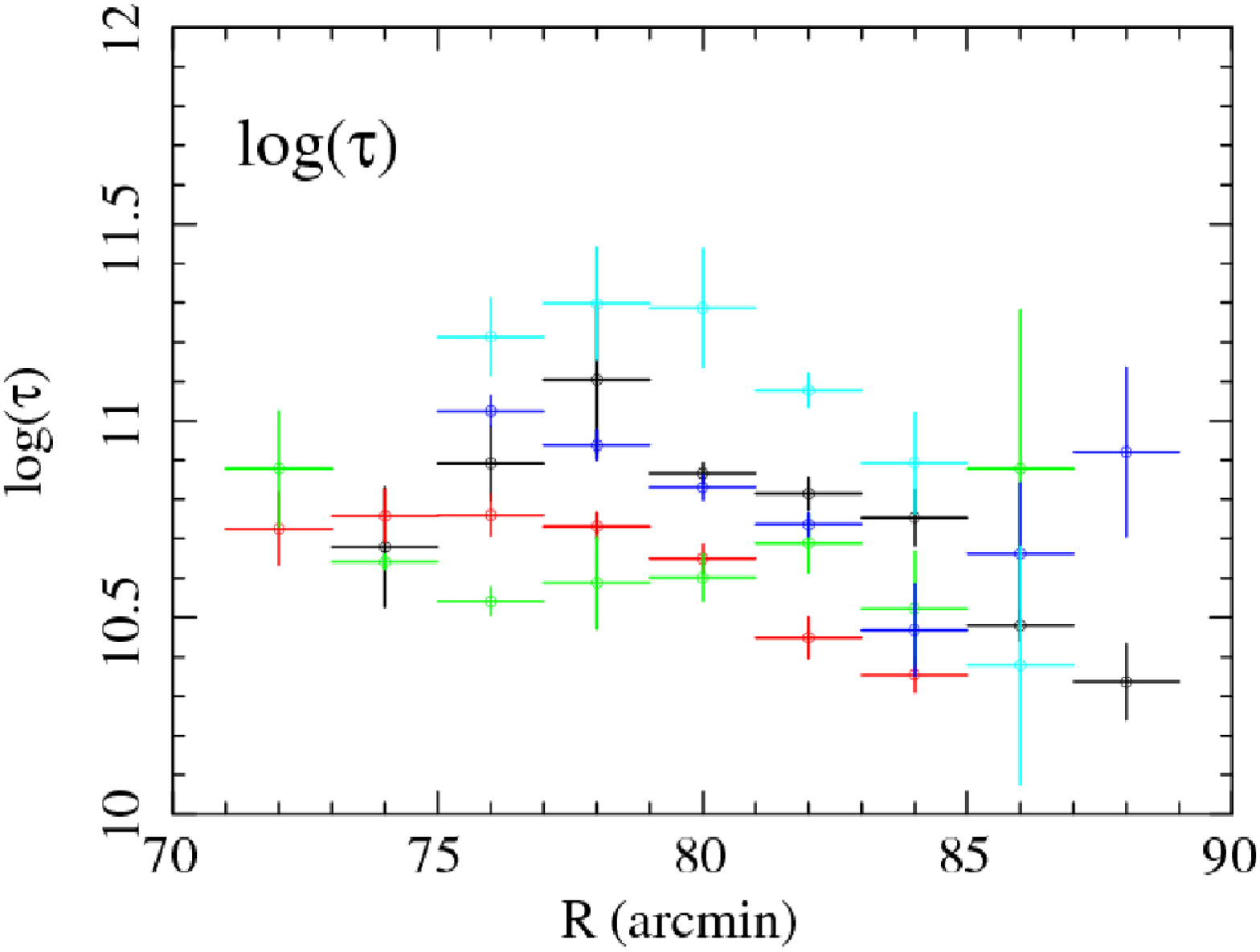}
    \FigureFile(60mm,0mm){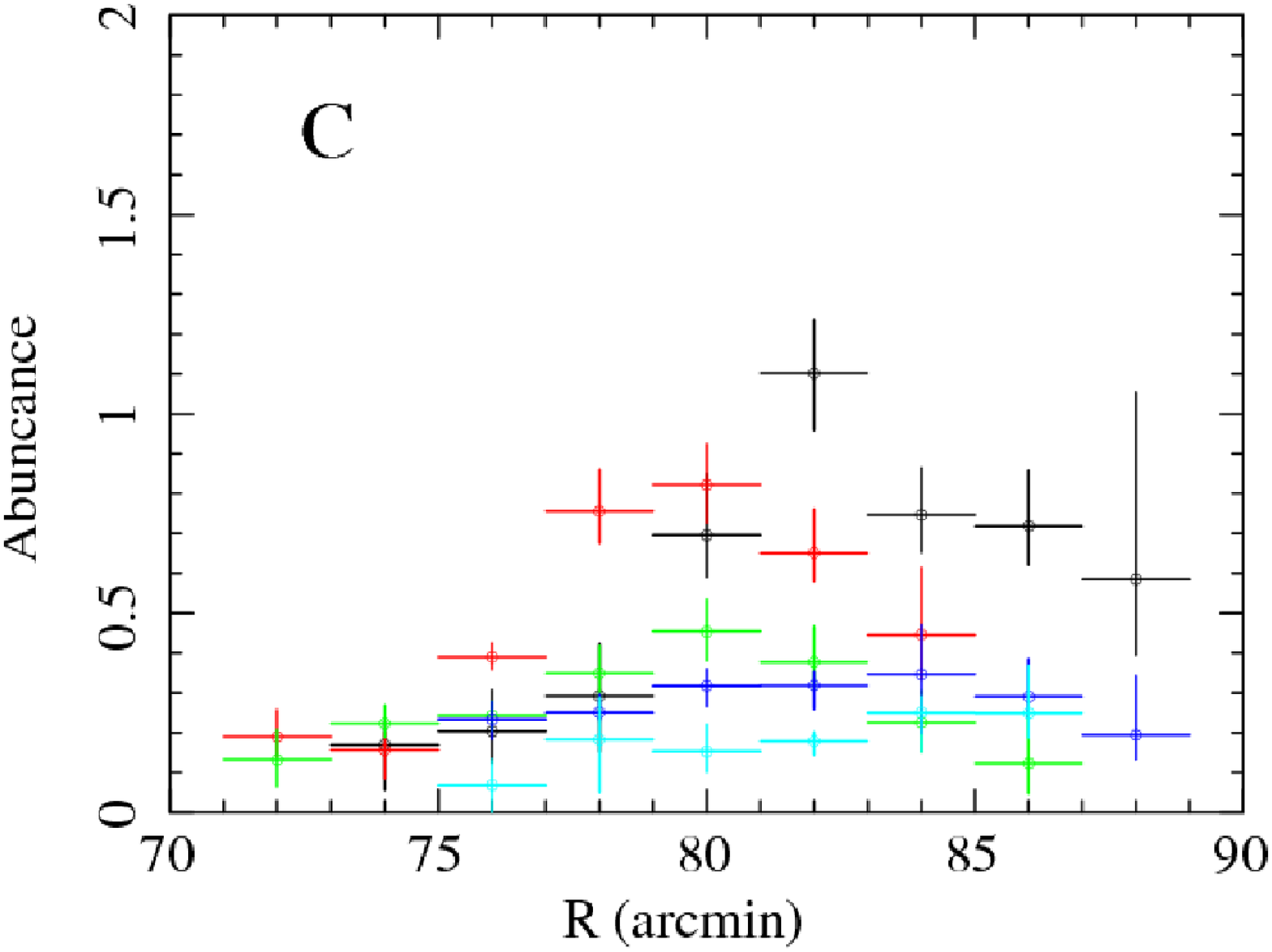}
    \FigureFile(60mm,0mm){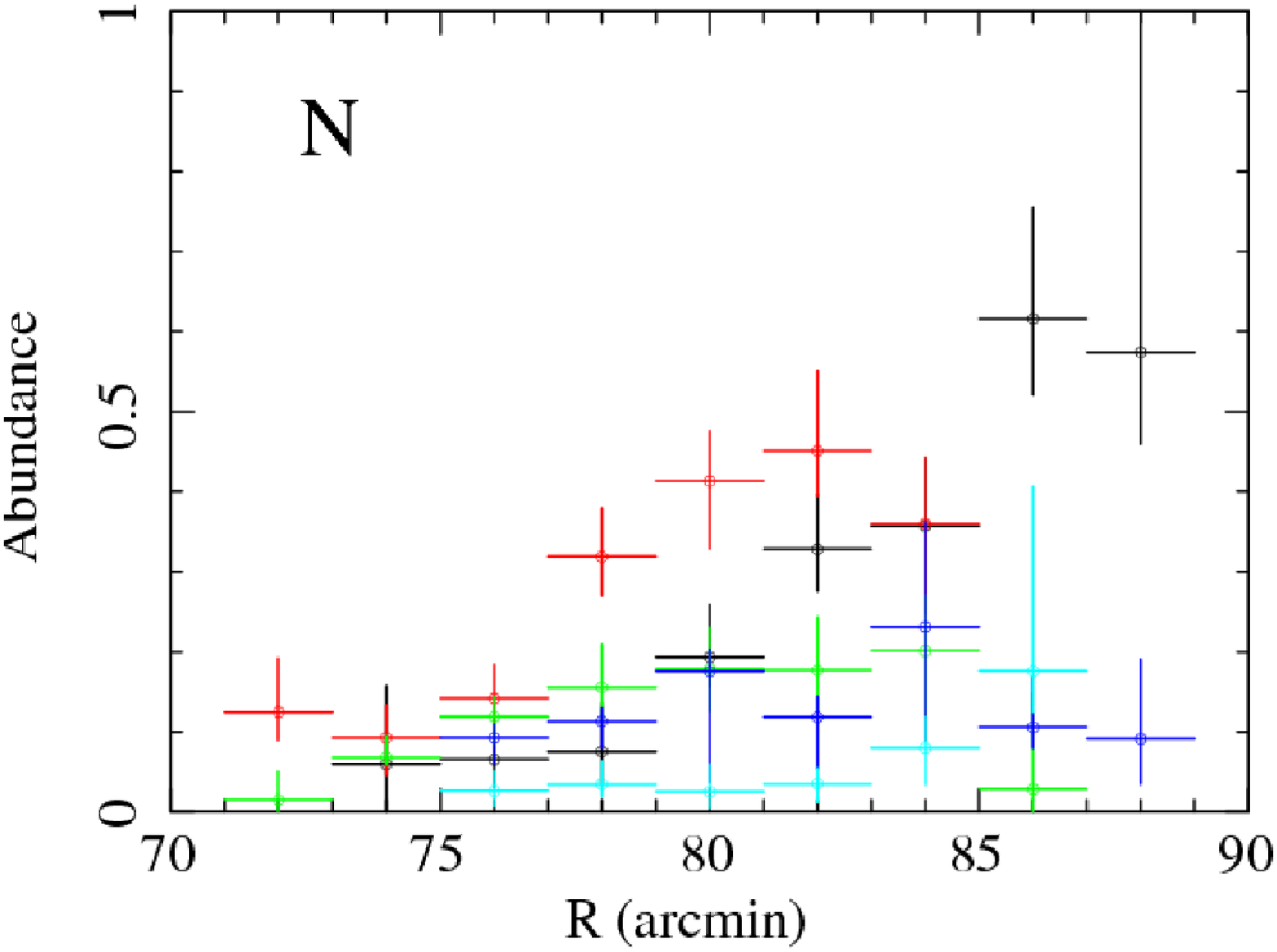}
    \FigureFile(60mm,0mm){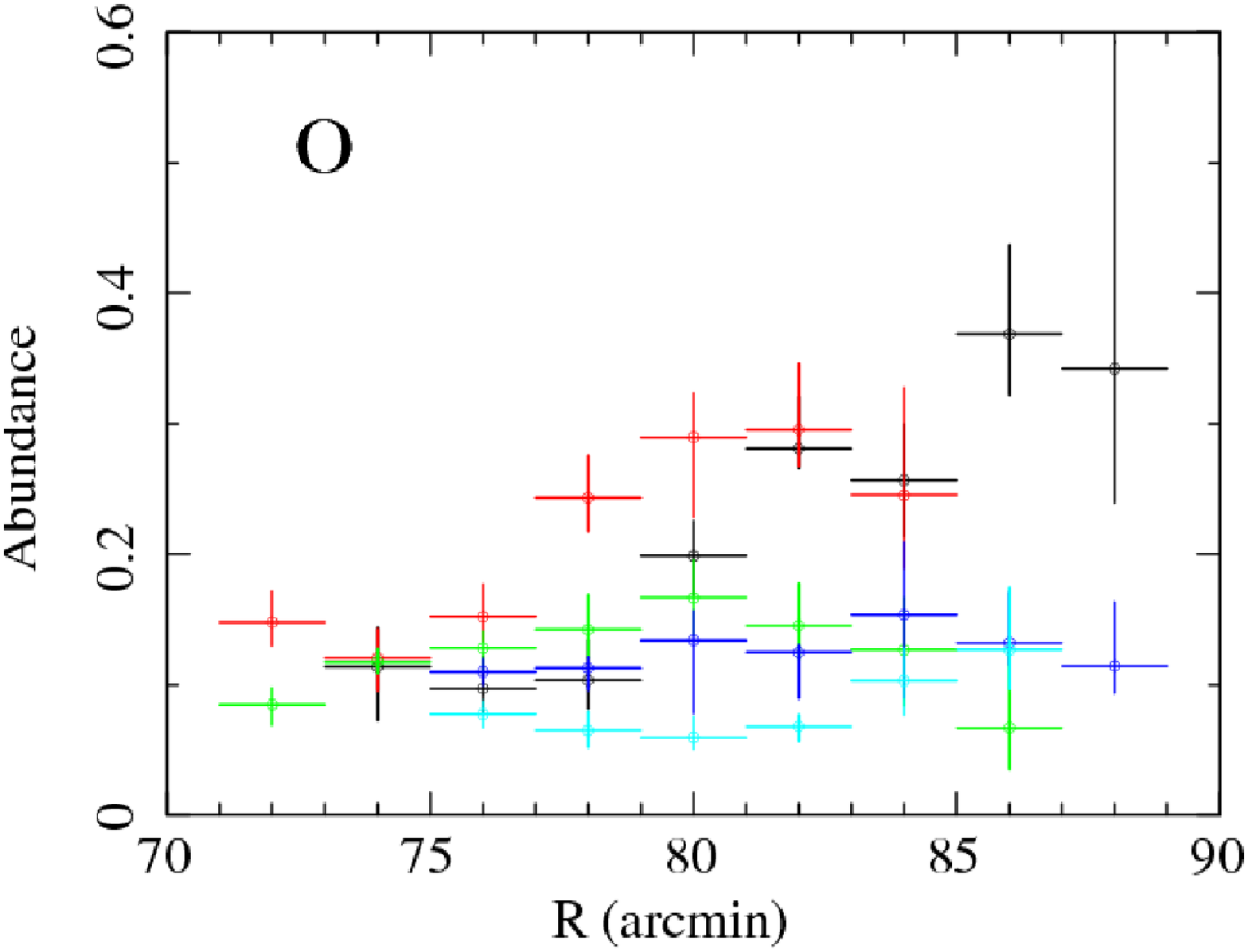}
    \FigureFile(60mm,0mm){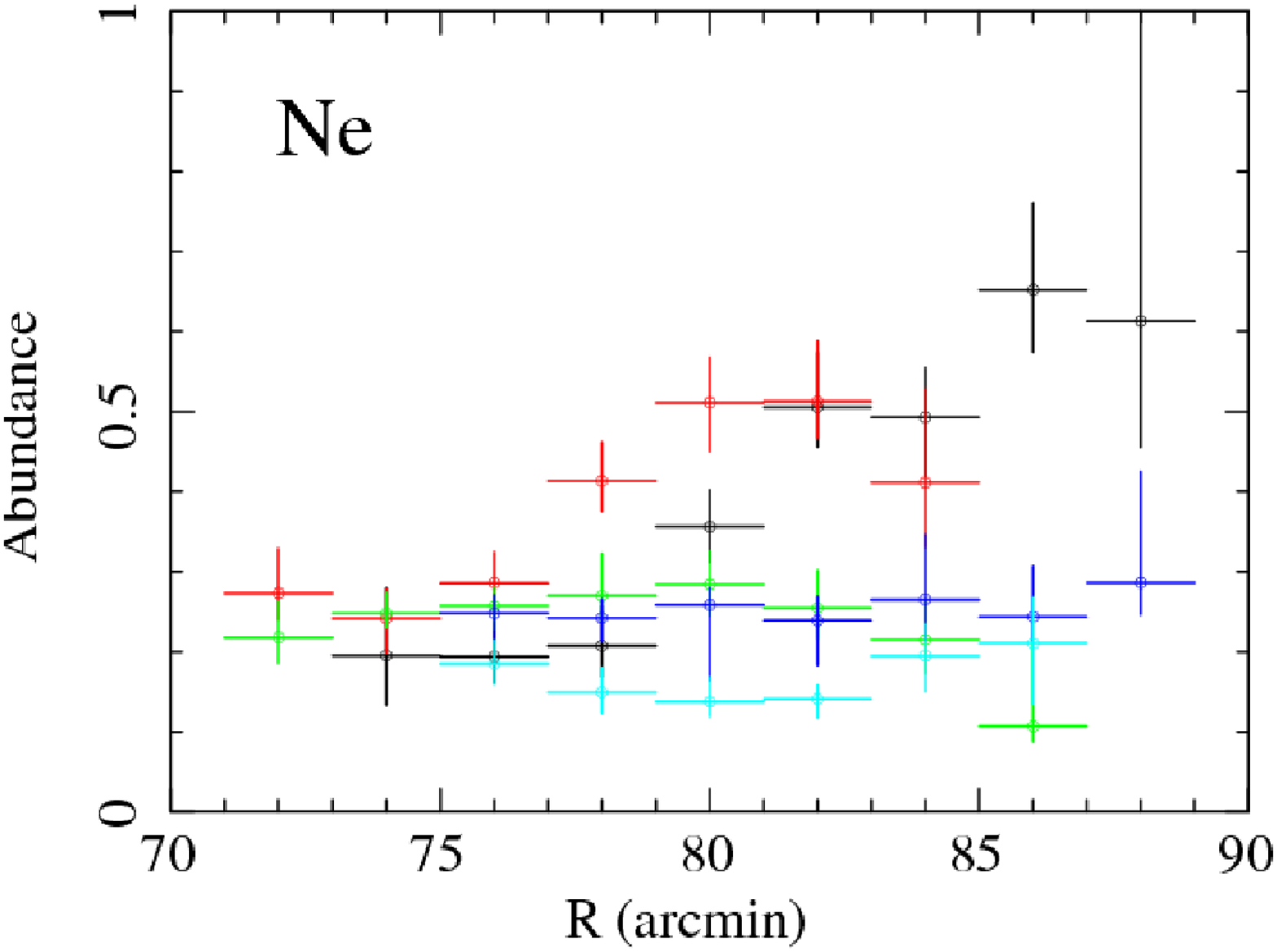}
    \FigureFile(60mm,0mm){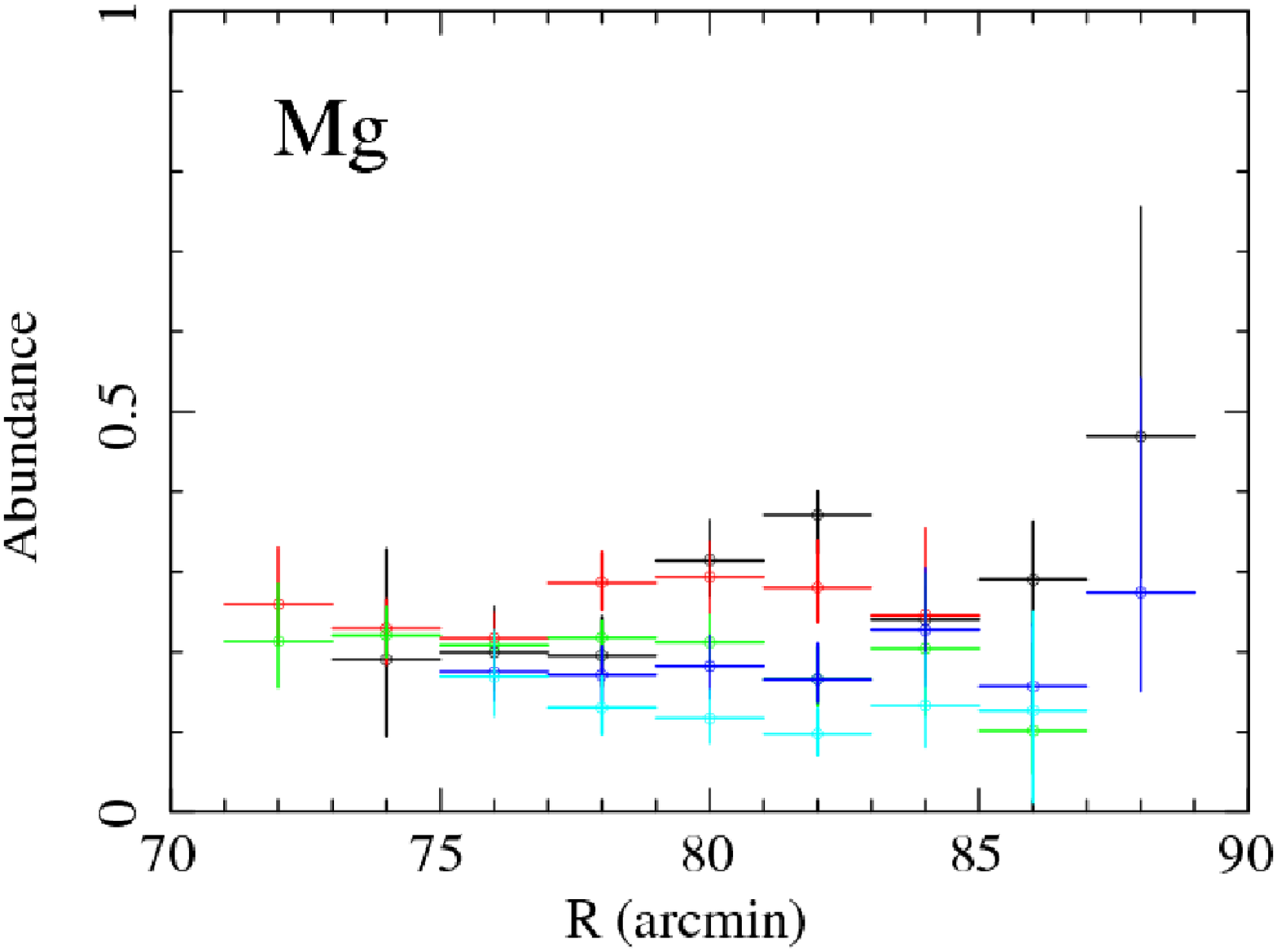}
    \FigureFile(60mm,0mm){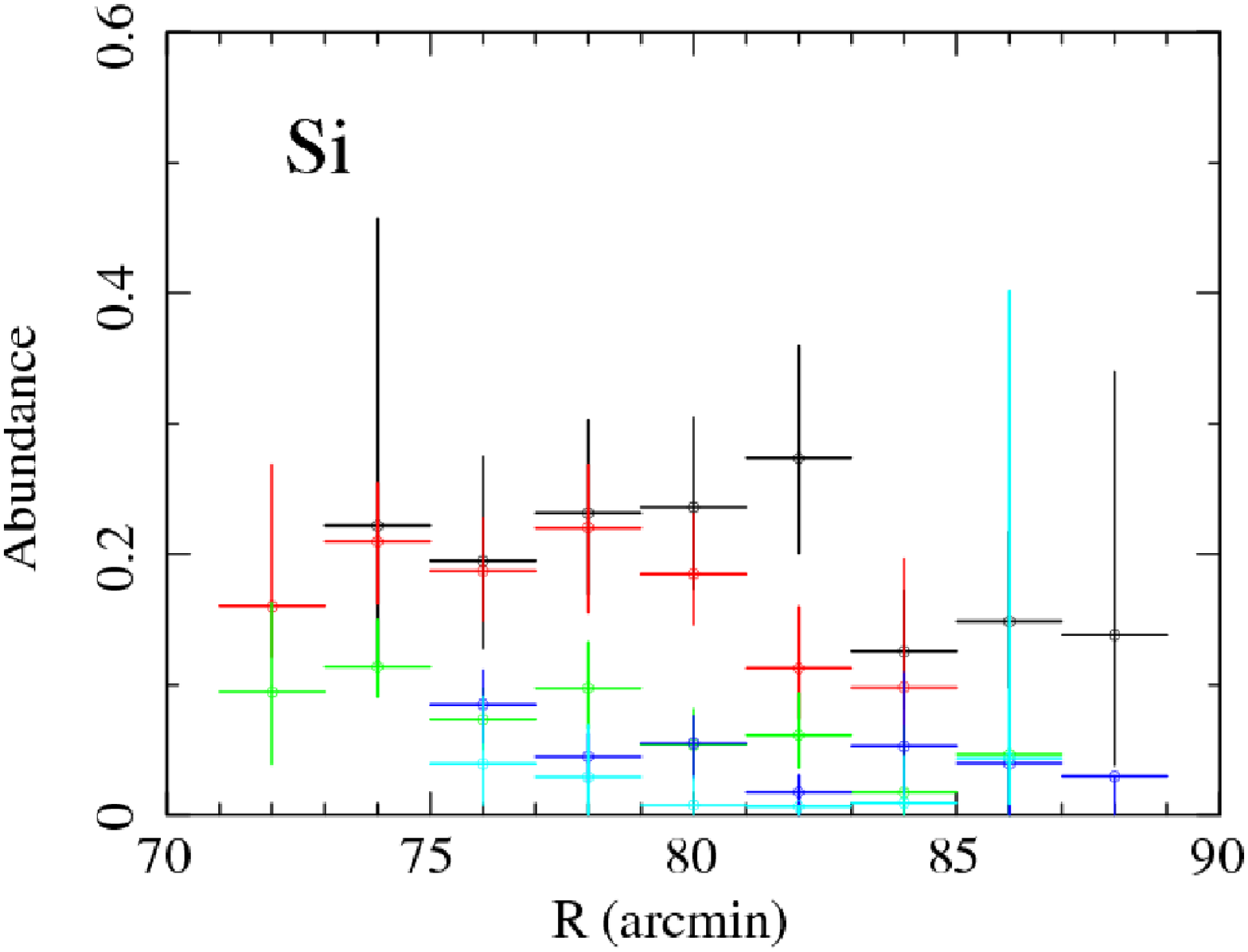}
    \FigureFile(60mm,0mm){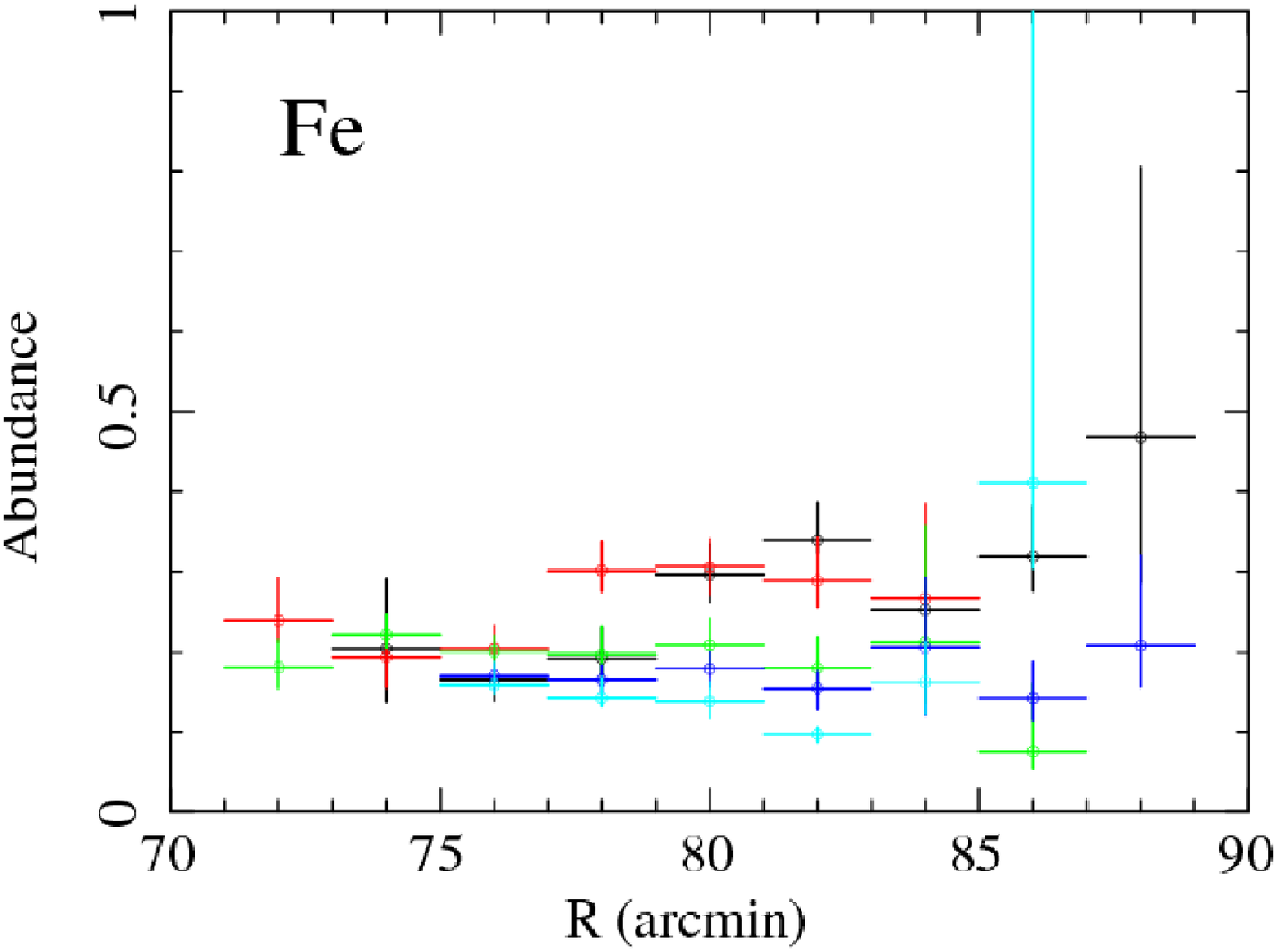}
  \end{center}
  \caption{Radial plot of $kT_e$, log($\tau$), and abundances of various elements as a function of $R$. Black, red, green, blue and light blue correspond to P21, P22, P23, P24 and P25, respectively.}\label{fig:funcr}
\end{figure}

\begin{figure}
  \begin{center}
    \FigureFile(70mm,0mm){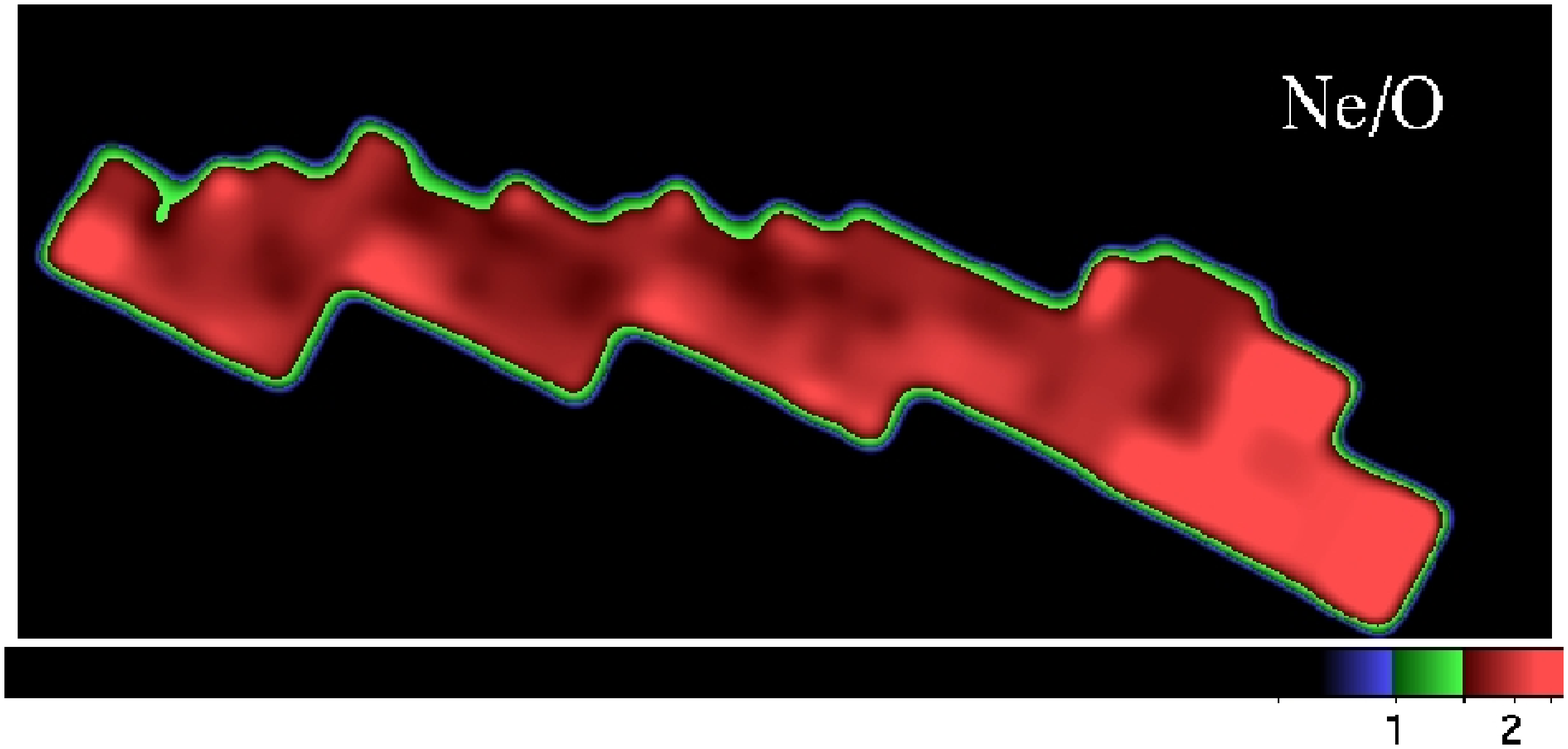}
    \FigureFile(70mm,0mm){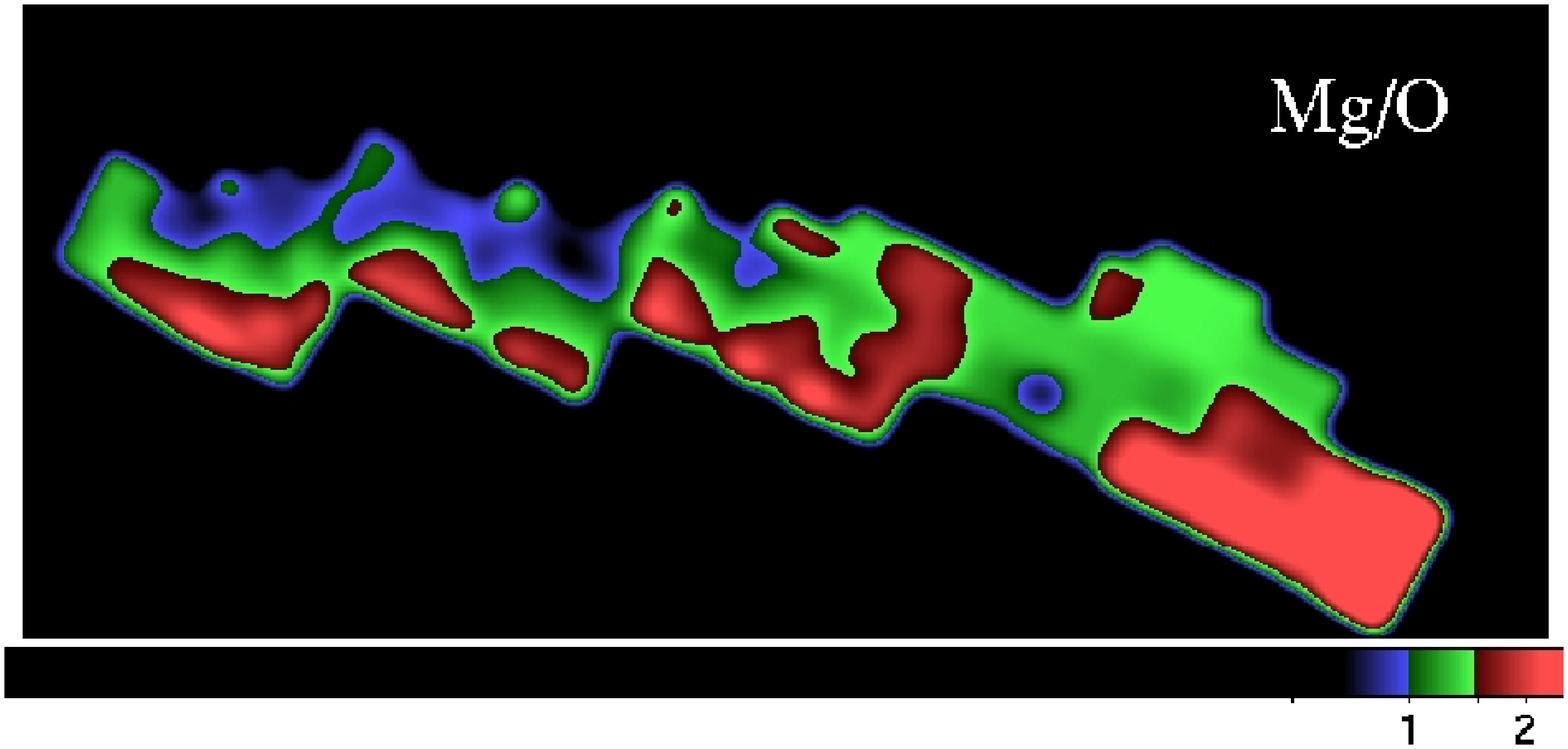}
    \FigureFile(70mm,0mm){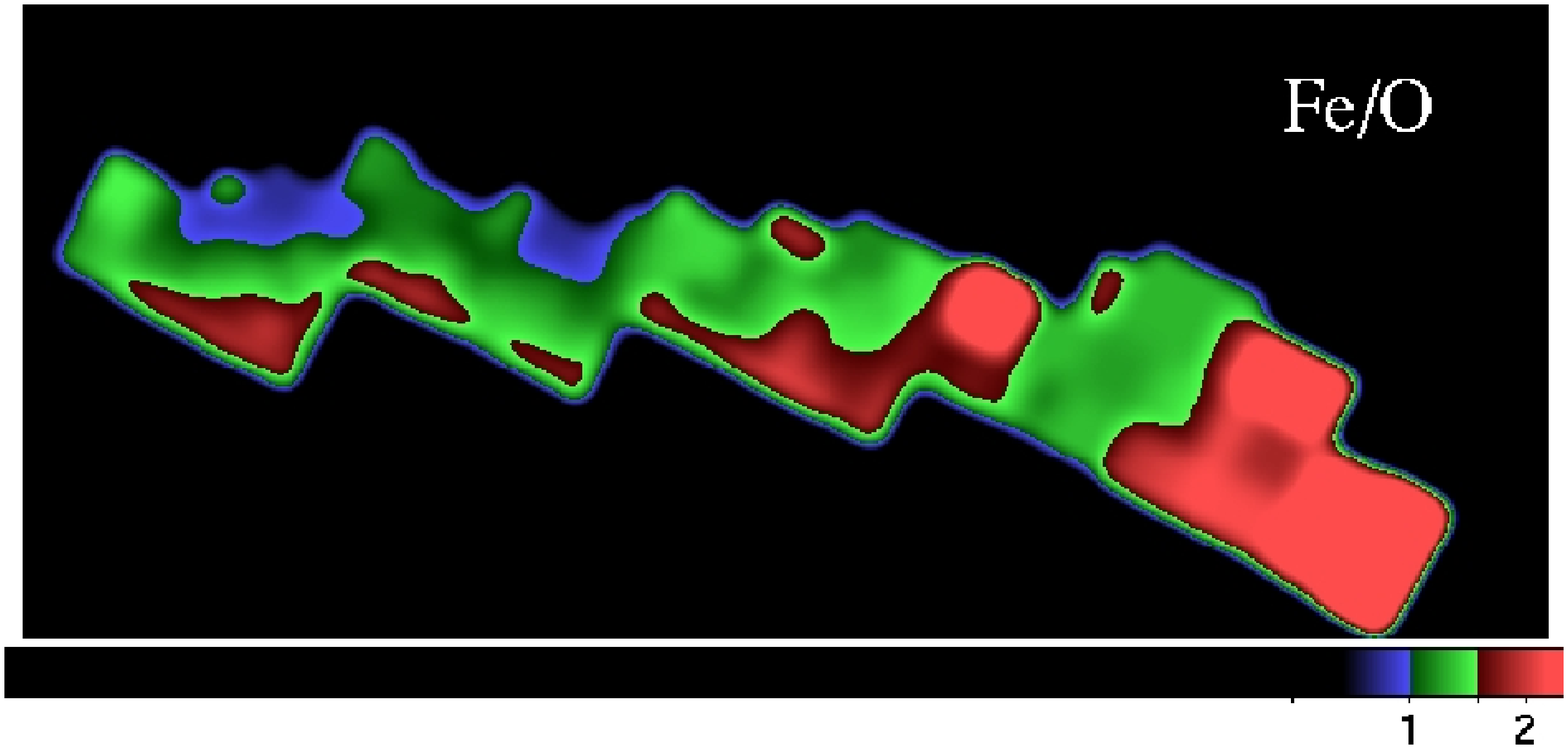}
  \end{center}
  \caption{Distributions of the relative abundances of Ne/O, Mg/O and Fe/O. Red, green and blue represent the regions where the relative abundance show more than 1.5, from 1 to 1.5, and less than 1, respectively.}\label{fig:ratio}
\end{figure}

Our FOV is next to that of \citet{Katsuda08NE}.
From figures \ref{fig:paramap} and \ref{fig:funcr}, we found that the abundances of N, O and Ne are relatively high at the outermost edge in P21-P22.
It continues into the abundance-enhanced region in NE3-NE4 and these values of abundances are similar to each other.
In addition, as shown by \citet{Katsuda08NE} in NE3-NE4, low abundance ratios ($<1$) of Mg/O and Fe/O are also seen at the abundance-enhanced region in our FOV.
Figure \ref{fig:ratio} shows the distributions of the ratios of Mg/O and Fe/O in our FOV.
For comparison, we also show the Ne/O map in the same figure.
Red, green and blue represent the regions where the relative abundances show more than 1.5, from 1 to 1.5, and less than 1, respectively.
From figure \ref{fig:ratio}, the values of the Ne/O are more than 1.5 times higher than the solar values for all regions.
On the contrary, the Mg/O and the Fe/O are both lower than the solar values at the abundance-enhanced regions in P21-P22.
In view of these facts, it is natural to consider that the abundance-enhanced regions in NE3-NE4 and P21-P22 have the similar origin. 

\citet{Katsuda08Chan} presumed that the abundance inhomogeneity in NE3-NE4 is due to the break of the cavity wall of the Loop. 
\citet{Falle82} and \citet{Shull91} both argue for an incomplete cavity of the Cygnus Loop, and the former also makes the case that the north part of the Loop is a very large scale breakout.
In that case, NE3-NE4 and our regions observed entirely should have normal ISM abundances.
However, \citet{Levenson97} and \citet{Levenson98} presented the spherical-cavity model and in that case, the blast wave where a small breakout exists could overrun the cavity wall and proceed first into the surrounding ISM with the ISM metallicity.
Thus, the abundances there could reflect the values of the ISM abundance.
Although the Cygnus Loop is almost circular in shape, some observations show that the thickness of the cavity wall is not uniform. 
\citet{Kimura09}\footnote{The data available at http://arxiv.org/pdf/0810.4704v1} observed northern region of the Loop from NE to SW with the \textit{Suzaku} satellite and revealed that the swept-up matter shell is very thin in just the west of center of the Loop. 
They estimated the diameter of this thin shell region to be $1^{\circ}$ and concluded that the breakout exists along the line of sight.
We can also see such breakout in the south of the Loop called the ``blow-out'' region \citep{Aschenbach99}.
\citet{Uyaniker02} and the other radio observations (\cite{Uyaniker04}; \cite{Sun06}) suggest the existence of a secondary SNR in the south. However, from the X-ray observation with \textit{XMM-Newton}, \citet{Uchida08} showed that the X-ray shell is thin in the blow-out region and concluded that the origin of the blow-out can be explained as a breakout into a lower density ISM.
These results suggest that the ambient density of the Loop is not uniform while the Loop seems to be spherical symmetry.
The abundance-enhanced regions previously observed have relatively weak surface brightnesses in X-ray (see figure \ref{fig:region}). 
As \citet{Levenson99} point out, the bright soft X-ray emission can result from a slower shock speed in denser cavity material. 
Conversely, the weak surface brightness at the abundance-enhanced regions observed suggest that the breakout or the thin cavity wall exists there.
 
\begin{figure}
  \begin{center}
    \FigureFile(80mm,0mm){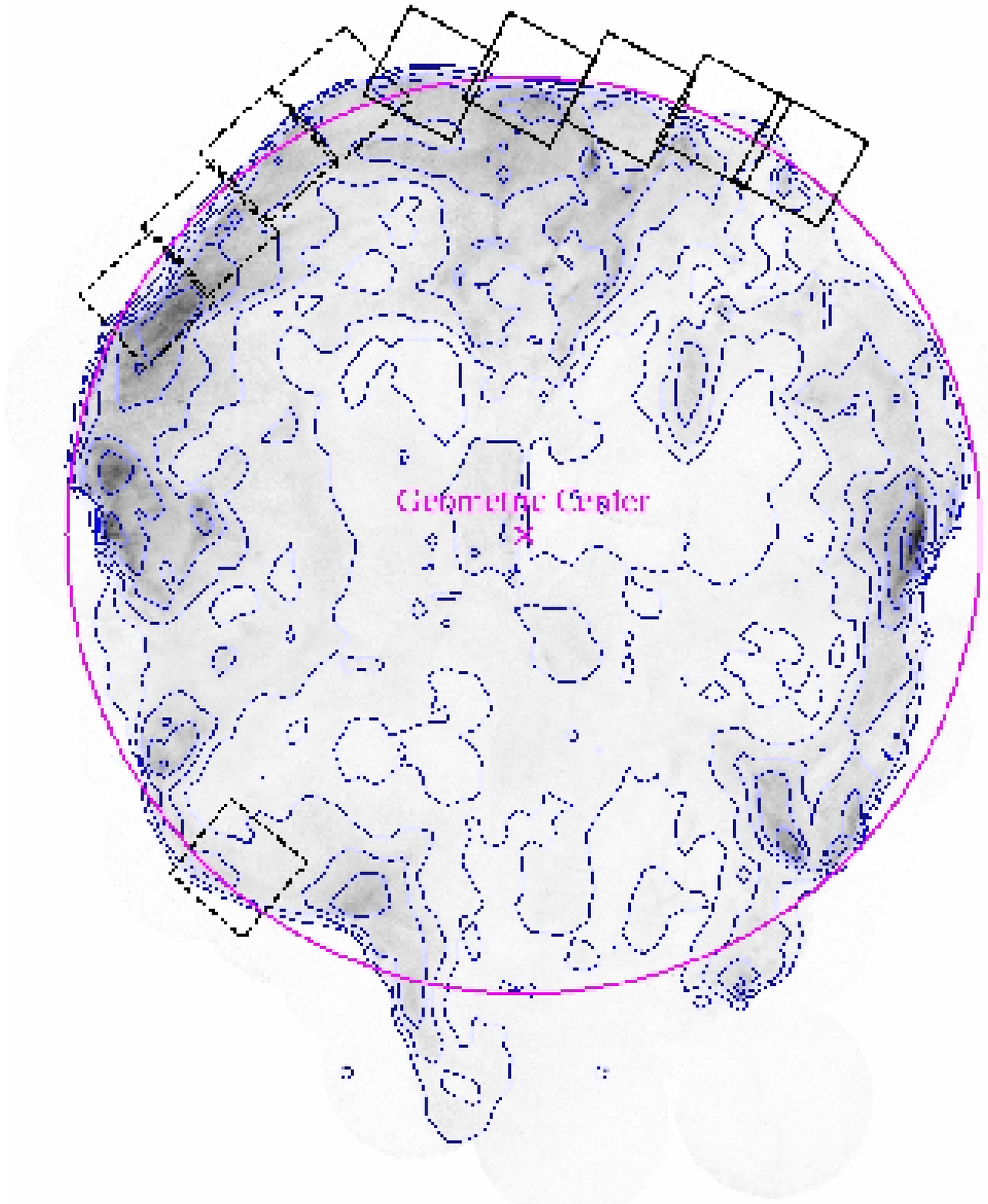}
  \end{center}
  \caption{\textit{ROSAT} HRI image of the entire Cygnus Loop and its contour (blue line) overlaid with our FOV (black solid line) and those of \citet{Katsuda08NE} and \citet{Tsunemi09} (dotted line). The geometric center and a circle with radius $\sim1^\circ.4$ are shown by the magenta lines.}\label{fig:cont}
\end{figure}

We also confirmed it from the morphological point of view.
Figure \ref{fig:cont} shows the \textit{ROSAT} HRI image of the entire Cygnus Loop and its contour (blue line).
The geometric center and a circle with radius $\sim1^\circ.4$ are shown by the magenta lines.
From figure \ref{fig:cont}, there are three outwardly-projecting shell regions where the rims are over the magenta circle, except the large south blow-out: 
a part of the NE rim (NE3 to P21), northwestern (NW) rim (P24 and P25), and the SE rim (P27).
Under the assumption that the supernova explosion occurred at the geometric center, the cavity walls toward these directions are considered to be thinner than those toward the other ones. 
Among these projecting shell regions, the NE and the SE rims clearly show the abundance-enhanced regions from previous (NE3-NE4, P27) and our observations (P21-P22).
It is reasonable to consider that the abundance inhomogeneities in these regions are derived from the breakout or the thinness of the cavity wall.
As for the NW rim, the abundances there show low values compared with those at the edges of the NE or the SE rims (see figure \ref{fig:paramap}).
However, even in this region, the abundances such as N , O and Ne at the outer edge are slightly higher than those at the inner region.
This fact suggests that the outer edge of the NW rim shows the sign of the abundance-enhancement as is the case with those of the NE and the SE rims.
Then, we concluded that the blast wave in the NW is now proceeding into the outside of the cavity wall and begins to interact with the surrounding ISM. 

We speculate that there are many breaks of the cavity wall in the Cygnus Loop where the abundances show the values of the ISM.
However, the spectra at these regions should be observed as the superpositions of those with the depleted abundances in the line of sight.
Therefore, the abundance-enhanced regions where we can observe are limited at the outer edges of the Loop.

\begin{table}
  \begin{center}
 \caption{Spectral fit parameters}\label{tab:spec2}
    \begin{tabular}{lccc}
       \hline 
      \hline
 Parameter & $R$=88${^\prime}$ (VNEI) & $R$=76${^\prime}$ (VNEI) \\
      \hline
      N$\rm _H$ [10$^{20}$cm$^{-2}$] & 1.3 $^{+0.7}_{-0.6}$ & 2.6 $^{+0.7}_{-1.0}$\\
     $kT_e$ [keV] & 0.34 $\pm$ 0.04  & 0.29 $^{+0.03}_{-0.02}$ \\
      C & 0.58 $^{+0.19}_{-0.47}$ & 0.29 $\pm$ 0.13 \\
       N & 0.57 $^{+0.52}_{-0.12}$ & 0.08 $^{+0.06}_{-0.04}$ \\
       O & 0.34 $^{+0.25}_{-0.10}$ & 0.10 $^{+0.03}_{-0.02}$ \\
       Ne & 0.61 $^{+0.40}_{-0.16}$ & 0.21 $^{+0.05}_{-0.04}$ \\
       Mg & 0.47 $^{+0.29}_{-0.18}$ & 0.20 $^{+0.05}_{-0.04}$ \\
       Si (=S) & 0.14 $^{+0.20}_{-0.10}$ & 0.23 $^{+0.07}_{-0.06}$ \\
       Fe (=Ni) & 0.47 $^{+0.34}_{-0.18}$ & 0.19 $^{+0.04}_{-0.03}$ \\
       log($\tau$) & 10.34 $^{+0.16}_{-0.09}$ & 11.10 $^{+ 0.10}_{-0.19}$ \\
       EM [10$^{20}$cm$^{-5}$] & 0.26 $\pm$ 0.01 & 3.05 $\pm$ 0.06 \\
$\chi ^2$/dof & 671/448 & 963/640 \\
      \hline
    \end{tabular}
 \end{center}
\end{table}

\section*{Acknowledgements}

This work is partly supported by a Grant-in-Aid for Scientific Research by the Ministry of Education, Culture, Sports, Science and Technology (16002004).  This study is also carried out as part of the 21st Century COE Program, \lq{\it Towards a new basic science: depth and synthesis}\rq. H.U. and S.K. are supported by JSPS Research Fellowship for Young Scientists.

\end{document}